\documentclass[10pt,a4paper]{article}
\usepackage[utf8]{inputenc}
\usepackage{latexsym,amsthm}
\usepackage{graphics,physics}
\usepackage{graphicx}
\usepackage{epsfig}
\usepackage{multirow}
\usepackage{amsmath}
\usepackage{hhline}
\usepackage{hyperref}
\usepackage{enumitem}
\usepackage{amsfonts}
\usepackage{xcolor}
\usepackage{soul}
\usepackage[font=small,labelfont=bf]{caption}
\usepackage[compat=1.1.0]{tikz-feynman}
\usepackage{subcaption}
\usepackage{hhline}
\usepackage{bm,ulem}
\usepackage{subcaption}
\usepackage{float}
\restylefloat{table}
\usepackage[margin=1in]{geometry}
\begin{document}

\newcommand{\old}[1]{{\textcolor{red}{\sout{#1}}}}
\newcommand{\new}[1]{{\textcolor{red}{#1}}}
\newcommand{\be}{\begin{equation}}
\newcommand{\ee}{\end{equation}}
\newcommand{\La}{\cal{L}}
\newcommand{\no}{ }
\newcommand{\ba}{\begin{eqnarray}}
\newcommand{\ea}{\end{eqnarray}}
\newcommand*{\pd}{\partial}
\newcommand*{\pdm}{\pd_{\mu}}
\newcommand*{\pdn}{\pd_{\nu}}
\newcommand*{\pdi}{\pd_{i}}
\newcommand*{\pda}[1]{\pd_#1}
\newcommand*{\bea}{\begin{eqnarray}}
\newcommand*{\eea}{\end{eqnarray}}
\newcommand*{\pref}[1]{(\ref{#1})}
\newcommand*{\img}{\mathrm{im}}
\newcommand*{\mn}{{\mu\nu}}
\newcommand*{\rr}{\mathbb{R}}
\newcommand*{\gh}{\mathrm{gh}}
\newcommand*{\sdet}{\mathrm{sdet}}
\newcommand*{\str}{\mathrm{str}}
\newcommand*{\uth}{^\mathrm{th}}
\newcommand*{\ust}{^\mathrm{st}}   
\newcommand*{\prefr}[2]{(\ref{#1}-\ref{#2})}
\newcommand*{\piclinecol}{red }
\newcommand{\D}{\displaystyle}
\setlength{\parskip}{4pt}

\newtheorem*{theorem}{Theorem}
\newtheorem*{definition}{Definition}

\title{\bf Unveiling Regions in multi-scale Feynman Integrals using Singularities and Power Geometry}

\author{\bf B. Ananthanarayan$^1$ \\ \bf Abhishek Pal$^2$ \\ \bf S. Ramanan$^3$ \\ \bf Ratan Sarkar$^1$} 

\date{%
    $^1 $Centre for High Energy Physics,\\ Indian Institute of Science,\\ Bangalore-560012, Karnataka, India \\ 
    \medskip \medskip \medskip \medskip
    $^2$Bartol Research Institute \& \\ Department of Physics and Astronomy, \\ University of Delaware, Newark, DE 19716, USA \\
    \medskip \medskip \medskip \medskip
    $^3$Department of Physics,\\ Indian Institute of Technology Madras,\\ Chennai-600036, Tamil Nadu, India }%

\maketitle

\begin{abstract}
We introduce a novel approach for solving the problem of identifying regions in the framework of Method of Regions by considering singularities and the associated Landau equations given a multi-scale Feynman diagram. These equations are then analyzed by an expansion in a small threshold parameter via the Power Geometry technique. This effectively leads to the analysis of Newton Polytopes which are evaluated using a Mathematica based convex hull program. Furthermore, the elements of the Gr\"{o}bner Basis of the Landau Equations give a family of transformations, which when applied, reveal regions like potential and Glauber. Several one-loop and two-loop examples are studied and benchmarked using our algorithm which we call ASPIRE. 
\end{abstract}

\section{Introduction}
\label{sintro}

Over the last couple of decades, Effective Field Theories (EFTs) have become mainstream in 
problems that involve separation of scales. While there are several applications in low-energy QCD, its
foray into high-energy processes is still relatively new and recent years have seen the
application of EFT ideas to problems in particle physics that involve several scales. Collider
physics provides a textbook example of a multi-scale problem involving particles with high energies
as well as comparatively low-mass particles such as protons. Multi-scale Feynman diagrams arise naturally in many branches of elementary particle physics. The analysis of such diagrams has led to new EFTs such as the soft-collinear effective theory, heavy quark effective theory and so on.  For an accessible introduction to the subject of effective field theories, see, e.g., refs.~\cite{Softbook,Scherer:2002tk, Wise:1997sg}. In a multi-loop problem, one typically encounters several mass 
and kinematic scales. A strategy that has been very effective in these
problems is the Method of Regions (MoR). This allows one to carry out asymptotic expansions of Feynman
integrals within dimensional regularization. The Feynman integral for a process is expanded as a sum
of simpler integrals which can then be done term by term. Further, the diagrams obtained in MoR can also be obtained from an appropriate EFT.
The different regions identified correspond to different EFTs characterized
by a threshold parameter.

An accessible example of separation of scales can be found in three flavor chiral perturbation theory, where one finds instances of diagrams
with a hierarchy of masses, namely the masses
of the pion, the kaon and the eta. Recent analytical progress can be found in refs.~\cite{Ananthanarayan:2018irl,Ananthanarayan:2017qmx}. The MoR was applied by Kaiser and Schweizer~\cite{KaiserSchweizer} for studying $\pi-k$ scattering processes in the context of ChPT. In this work, we also study this process within our framework to identify the associated regions.

The application of the MoR to multi-scale problems has been studied now for nearly two decades, starting with the fundamental work of Beneke and Smirnov~\cite{beneke_smirnov_1997}. Subsequently, this approach has been used for Drell-Yan Processes~\cite{Bonocore:2014wua}, studying massless fermionic processes at next to leading order~\cite{masslessFermionic2007}, investigating processes in the Sudakov limit~\cite{Jantzen:2006jv} and so on. To find the leading order contribution of Feynman integrals at threshold, one needs to sum contributions from several regions which span the entire loop momentum space. The contributing regions result from the presence of a hierarchy of masses, or from components of some momenta becoming small or large compared to others. 

In a recent work, Pak and Smirnov~\cite{pak_smirnov} proposed a geometric algorithm which could be automated to find the regions of a given Feynman diagram. By going to the Alpha-representation, they show that the problem can be turned into a geometric one and can be solved by finding the convex hull of a set of carefully chosen points. The
Mathematica package developed that automates the finding of regions, called asy.m, which we will 
refer to as ASY, uses a C++ -based QuickHull algorithm ~\cite{cpp_qhull}.

In its original implementation, ASY fails to identify potential and Glauber regions. These
regions usually manifest as differences between the Alpha-parameters. In the updated asy2.1~\cite{Jantzen:2012mw,asy_code} a new feature called \emph{PreResolve} was introduced that  eliminates the differences of Alpha-parameters from the Symanzik polynomials using linear transformations. We will refer to the upgraded version of the code as ASY2. 

In the present work, we approach the problem from another perspective by looking at the singular structure of the Feynman integral in the Alpha-representation. The analysis of the Pinched Singular Surfaces, in the momentum space, in connection to the regions~\cite{Eden:1966dnq,Libby:1978qf,Libby:1978bx,Collins:2011zzd} is well understood in terms of the Landau Equations. We set up and study the set of Landau Equations, in the alpha parameter space, for a given
process using the Gr\"{o}bner Basis and derive a criterion for the determination of the transformations 
required to reveal the regions within the framework of Power 
Geometry~\cite{bruno2015,bruno2012,bruno:book}, thus, demonstrating a new way of solving the 
problem of finding regions.

The paper is organized as follows. In section~\ref{sect:form}, we introduce elements essential to our algorithm. We review the strategy of the MoR in sub-section ~\ref{sect:mor} and then give an overview of the geometric approach discussed by Pak and Smirnov 
in~\ref{sect:asyimplementation}, by considering an example from~\cite{pak_smirnov}.  
In~\ref{sect:pss}, we revisit the problem of finding the regions through an 
alternative approach that links the analytic structure of the Feynman integral to the regions.  In fact, the contents of \ref{sect:pss}
constitute the important theoretical progress being reported in
this work. We conclude the section~\ref{sect:form} by summarizing the
steps of our algorithm, ``Algebro-geometric analysis of Singular Polynomials for Identification of REgions (ASPIRE)".  We apply the algorithm to one-loop
and two-loop examples in section~\ref{sect:examples}. Finally, in section~\ref{sect:disc-conclusions}, 
we discuss our results and present our future goals. The details of our Mathematica notebooks and external packages used in this work are 
given in the appendix.  

\section{Formalism}
\label{sect:form}
In this section, we set up the formalism for identifying the different regions using the singular
structure of the Feynman integral. This process can be automated using ideas from
power geometry. However, for the sake of completeness, we also summarize the technique of Pak and  Smirnov in a subsequent sub-section.

\subsection{Method of Regions}
\label{sect:mor}

The technique of the MoR was proposed in an attempt
to analytically approximate various processes within perturbation 
theory~\cite{beneke_smirnov_1997,Jantzen:2011nz,Smirnov:1997ct,Smirnov_applied}. The idea of the 
MoR is to provide an expansion of the integrand in ratio of the scales 
involved, usually in the form of low-energy scale to high-energy scale. This results in expressing
the original Feynman integral as a sum over simpler integrals, all of which need to be integrated 
over their corresponding domains, which are called regions. 

Let us consider an example of a process 
from~\cite{pak_smirnov} to illustrate the idea.

\begin{figure}[h]
    \begin{center}
    \begin{tikzpicture}
        \begin{feynman}
        \vertex (a);
        \vertex[right=1.5cm of a] (b);
        \vertex[right=2cm of b] (c);
        \vertex[right=1.5cm of c] (d);
     \diagram[small, horizontal=b to c]
     {
        (a) --[fermion, edge label=\(q\)] (b),
        (c) --[fermion,edge label=\(q\)] (d),
        (c) --[fermion, half right, edge label=\(k\)] (b),
        (b) --[fermion, half right,edge label=\(q+k\)] (c),
        };
        \end{feynman}
    \end{tikzpicture}    
    \end{center}
\caption{Self energy diagram for a scalar field with mass $m$}
\label{fig:pol}
\end{figure}
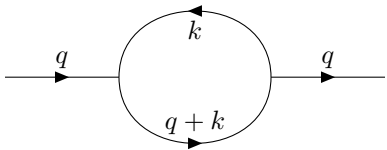

Fig.~\ref{fig:pol} shows a one-loop process given by
\begin{equation}
I(q^2, m^2) = \int \frac{d^d k}{(2\pi)^d} \frac{1}{(k^2 + m^2) ((k+q)^2 + m^2)}
\label{eq:pol}
\end{equation}
Let us consider the integral in Eq.~\ref{eq:pol} in the limit when
$|q^2| \gg m^2$ or $\rho \equiv |m^2/q^2| \ll 1$. Note that the loop momentum spans over all values ranging from $-\infty$ to $\infty$, thereby ruling out a naive Taylor expansion of the integrand. Let us denote the two denominators appearing in Eq.~\ref{eq:pol} as 
$D_1 = k^2 + m^2$ and $D_2 = (k + q)^2 + m^2$. As discussed in~\cite{pak_smirnov}, the following regions become relevant:
\begin{enumerate}
\item $|k^2| \sim |q^2|$: Here $D_1 = k^2$ and $D_2 = (k + q)^2$.
\item $|k^2| \sim m^2$: Then $D_1 = k^2 + m^2$ and $D_2 = q^2$.
\item $|(k + q)^2| \sim m^2$: This results in $D_1 = q^2$ and $D_2 = (k+q)^2 + m^2$.
\end{enumerate}

The original integral is the sum of the contributions of the above three regions each of which is evaluated within dimensional regularization. While it is easy to identify the regions in this particular example by looking
at the different scales in the Feynman integral, the procedure is a non-trivial task
in general, especially when one wants to evaluate multi-loop processes. Even at the one-loop
level, one encounters non-trivial regions, that involve a
multitude of scales. Another obvious difficulty in identifying regions is when the components of the momenta scale differently or when scalings of the difference of momenta are involved.
In the following, we review a specific implementation by Pak and Smirnov that
allows to isolate the regions.
 
\subsection{ASY Implementation}\label{sect:asyimplementation} 
Pak and Smirnov proposed an algorithm to automate 
the process of finding regions, which was documented in ~\cite{pak_smirnov,Jantzen:2012mw} 
together with their codes  ``asy.m'' and ``asy2.m'' (referred to here as ASY and ASY2). The 
second version adds crucial improvement to the first, as we 
summarize in the ensuing discussions. The basic idea is to parameterize the Feynman integral using 
the Alpha-parameters and then carry out the integration over the loop momenta using dimensional regularization to obtain its Alpha-representation. Expanding a process in the momentum space in scalings of momenta (or its components) is equivalent to expanding the Alpha-representation in scalings of combinations of Alpha-parameters.
 
ASY starts with expressing the integral in the Alpha-representation. For the process in Fig.~\ref{fig:pol} this yields,
\begin{equation}
	I(q^2, m^2) = \Gamma(2 - d/2)\int_0^\infty \int_0^\infty dx_1\, dx_2\, \delta(1 - x_1 - x_2) \,{\cal U}^{2 - d}\,{\cal F}^{d/2 - 2},
  \label{eq:alpha-rep-I1}
\end{equation}
where ${\cal U}$ and ${\cal F}$ are the Symanzik polynomials given by:
\begin{equation}
 {\cal{U}} = x_1 + x_2,
 \label{eq:U}
\end{equation}
and
\begin{equation}
 {\cal{F}} = x_1 x_2 (q^2 + 2 m^2) + x_1^2 m^2 + x_2^2 m^2,
 \label{eq:F}
\end{equation}
which are homogeneous in the Alpha-parameters, $x_1$ and $x_2$. Furthermore, Pak and 
Smirnov, build a new polynomial ${\cal U} \cdot {\cal F}$ that allows for a combined analysis of 
both the polynomials. All the terms in the leading order Symanzik polynomials have the same scaling 
in terms of the expansion (or threshold) parameter, $\rho \equiv \vert m^2/p^2 \vert$. More 
precisely, if one constructs a set of vector exponents $\lbrace r_i \rbrace$ for each monomial and 
scalings $\lbrace v_i \rbrace$ such that $x_i$  scales as $\rho^{v_i}$ then $x_i^{r_i}$ will scale as 
$\rho^{r_i v_i}$. Hence, the monomials in the Symanzik polynomials will scale as

\begin{equation}
\prod_i \rho^{r_0 v_0}\rho^{r_i v_i},
\end{equation}
where $r_0$ is the exponent of the threshold parameter appearing in the prefactor of the monomials and $v_0=1$. We can now construct an $n+1$ dimensional vector with components $\bm{r}=(r_0, r_1, \cdots r_n)$ such that the vector exponents obtained from each monomial of the leading order expansion, $v_{\rm leading}$, lie on a plane described by the equation
\begin{equation}\label{Eq:cond1}
\bm{r} \cdot \bm{v}_{\rm leading} = c,
\end{equation}
where $c$ is a constant.
All the other terms which do not appear at leading order, i.e $v_{\rm subleading}$, will lie above the surface described by
eq.(~\ref{Eq:cond1}). These points satisfy the condition
\begin{equation}
\bm{r} \cdot \bm{v}_{\rm subleading} > c.
\end{equation}

In general, it can be seen that if we construct the collection of vector exponents with $n+1$ 
components and plot them in an $n$-dimensional sub-space then the leading order terms
corresponding to a region will be points lying on the same surface and all the other points will lie above 
it. This immediately leads to the interpretation of the surface as a bottom facet of the convex 
hull of the set of vector exponents. Thus, finding the regions amounts to finding the convex hull of the 
set of vector exponents and then finding the normals of the lower facets of the convex hull.

For the Symanzik Polynomials in Eqs.~\ref{eq:U} and~\ref{eq:F}, ASY first calculates the product
\begin{equation}
{\cal U} \cdot {\cal F} = m^2 x_1^3+3 m^2 x_1^2 x_2 + 3 m^2 x_1 x_2^2 + m^2 x_2^3 + q^2 x_1^2 x_2 + q^2 x_1 x_2^2 
\label{eq:UF}
\end{equation}
from which one can extract the set of vector exponents(using threshold parameter $\rho \equiv |m^2/q^2|$)   
\[r=\lbrace (1,3,0),(1,2,1),(1,1,2),(1,0,3),(0,2,1),(0,1,2) \rbrace\]
 
\begin{figure}
\centering
\includegraphics[scale=0.7]{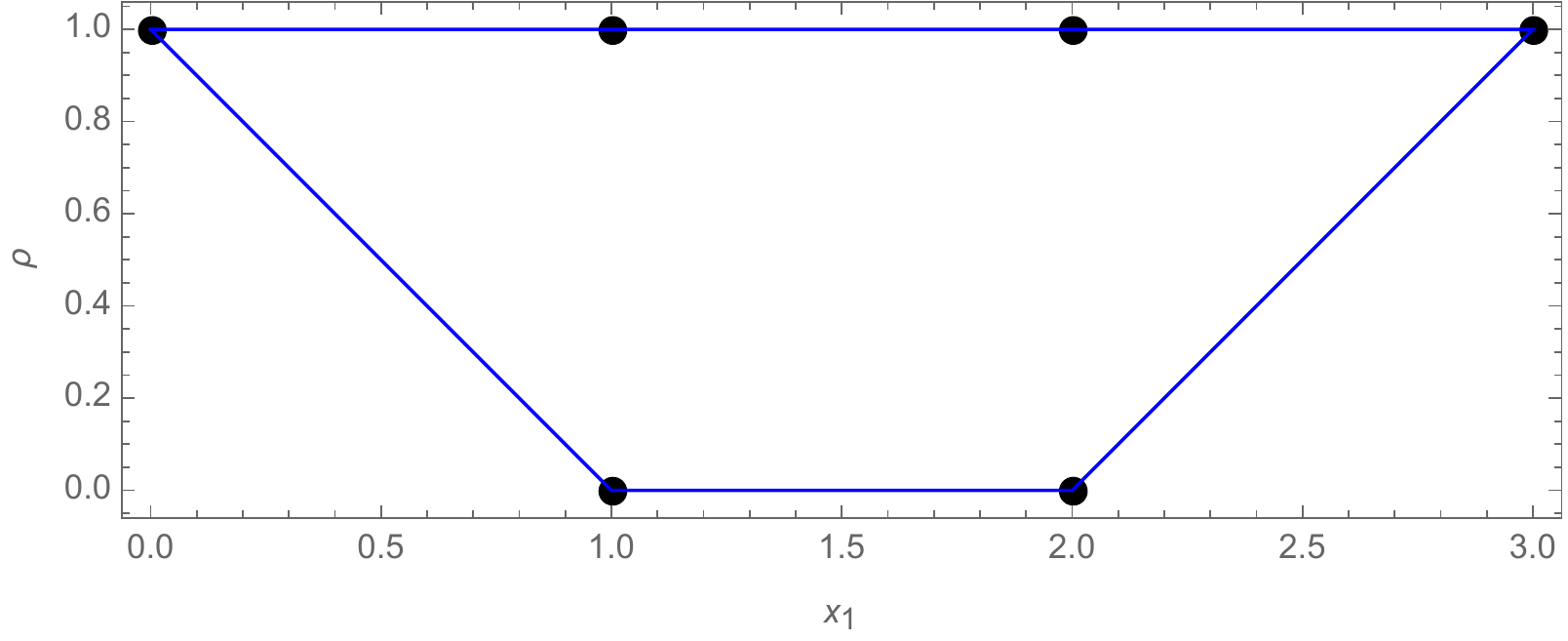}
\caption{Convex Hull of the Projected set of vector exponents}
\label{fig:Asy-conv-hull}
\end{figure}

Each vector exponent, corresponds to a monomial in Eq.~\ref{eq:UF} and the
components give the exponents of the expansion parameter $\rho$, followed by the set 
$\lbrace\,x_i\,\rbrace$. ASY projects this set of vector exponents onto a lower dimensional subspace. 
The new set of exponents, after performing the projection, is now $r=\lbrace (1,3),(1,2),(0,2),(1,1),(0,1),(1,0) \rbrace$. Following the 
discussion earlier in this section, ASY finds the convex hull of this 
projected set of exponents as shown in Fig.~\ref{fig:Asy-conv-hull} and then finds the normals of the bottom facets of the convex hull.

For example, the leading order term for the hard region (defined by $x_1 \sim \rho^0, x_2 \sim \rho^0$), is $q^2 x_1^2 x_2 + q^2 x_1 x_2^2$ corresponding to the projected points $(0,2), (0,1)$ lie on the plane with the normal vector (1,0). This is also seen from the fact that the leading order terms are independent of $\rho$ and thus scale as $\rho^0$. All the other terms have scalings larger than $\rho^0$ and thus will lie above the plane.

The first version of the code ASY could identify regions, except the Glauber and the potential, in several instances. The second version of the code
ASY2 fixes this shortcoming by linearly transforming the Alpha-parameters, eliminating any term in 
${\cal U} \cdot {\cal F}$ that appears with a difference between the Alpha-parameters. This is done 
by an option \emph{PreResolve} in the code. 
The main feature that distinguishes the two versions of the
ASY codes are the implementation of linear transformations
which allows the code to now identify the potential and
Glauber regions. This approach has been very successful in determining
regions and has been applied to several examples. 
To this extent, ASY also provides a very useful crosscheck
on prior studies besides a useful benchmark for
comparison.

In the following sub-sections we will demonstrate a new way of solving the problem of finding regions based on the singularities of Feynman integral in Alpha representation.
\subsection{Determination of the regions using the analytic structure of the propagator} 
\label{sect:pss}

The different regions, where a particular mass or kinematical scale becomes important can be linked
to the underlying singularities of the Feynman integral. In the following, we will introduce the main concepts 
and motivate the ideas that will lead to the development of the final algorithm. We first give an overview of 
the singularities that are of interest for our problem, followed by a review of the basic understanding of 
particle thresholds as \emph{pinched singularities} in momentum space. This interpretation is well understood and 
can be mathematically expressed using a set of equations called the Landau equations. 
Since expansions in the neighborhood of the singular surfaces give us the leading order 
behavior of Feynman amplitudes, we perform similar expansions in the Alpha-parameter space in 
carefully chosen neighborhoods of the singular points. This requires us to use techniques from the field 
of power geometry. We then motivate the use of Gr\"{o}bner basis for the identification of all 
neighborhoods of the singular points.

\subsubsection{Singularities and Threshold processes}
\label{sect:pss-sing}

\begin{figure}[ht]
\centering
\includegraphics[scale=0.5]{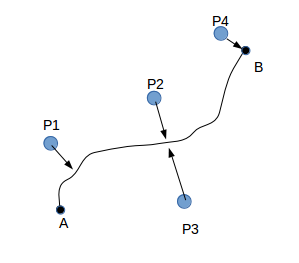}
\caption{Types of singularities: P1 is a simple pole, P2 and P3 are \emph{Pinched Singularities} and P4 is an
\emph{End Point Singularity}. While the contour between the points $A$ and $B$ can be deformed so as to avoid
the simple pole $P1$, the same is not true for the pinch and end-point singularities.}
\label{Fig:Sing}
\end{figure}

Understanding the analytic structure of the amplitude is crucial to identifying the 
different regions. The poles in the integrand of the amplitude for a given process are functions of kinematical
invariants, loop momenta etc. Therefore, when these parameters vary, the poles in the integration domain 
move. In the case of isolated singularities, it is always possible to deform the contour of integration to avoid
these singular points. However, sometimes, the poles migrate so as to pinch the contour of integration
(pinch singularities) or move to one of the end point of the integration (end point singularities) as 
illustrated in Fig.~\ref{Fig:Sing}. In such cases, these singularities cannot be avoided by contour 
deformations. 

The condition for a point to be one of these unavoidable singular points is the usual condition for establishing a singularity for a 
polynomial. For an arbitrary polynomial $g(\lbrace\,x_i\,\rbrace)$ that appears in the denominator 
of the Feynman integral, the point $x_i$ is singular point iff 
\begin{eqnarray}
\nonumber
 g(\lbrace x_i \rbrace) = 0, \\ 
\frac{\partial\mathit{g}}{\partial x_i} = 0. 
\label{eq:poly sing}
\end{eqnarray}

 Therefore, at these unavoidable singular points, hereby referred to as just singular points, the integrand diverges. We will now adopt the approach of Norton and 
Coleman~\cite{NortonColeman:1965xm}, also discussed in~\cite{Sterman:1994ce}, to explain the connection between the 
singularities and physical events.

Consider a general Feynman amplitude in the Alpha representation
\begin{equation}
I = \int \prod_i d^dk_i \prod_j d\alpha_j \delta\left(\sum_j \alpha_j -1\right)\mathit{f}(\lbrace q_j \rbrace)\ D^{-n}
\label{eq:feyn}
\end{equation}
  where,
\begin{equation}
\label{eq:D pol}
D = \sum_j \alpha_j(q_j^2-m_j^2)
\end{equation}
and $k_i$ are the loop momenta, $q_j$ are internal momenta which are linear functions in loop and external momenta, $m_j$ are the masses and $\alpha_j$ are the Alpha-parameters.

For a singularity, corresponding to Eq.~\ref{eq:poly sing} we have the conditions:

\begin{equation}
q_j^2 = m_j^2 \,\, \text{or} \,\ \alpha_j = 0,
\label{eq:landau1}
\end{equation}
and
\begin{equation}
 \frac{\partial}{\partial k_i}\sum_j \alpha_j(q_j^2-m_j^2) = 0.
 \label{eq:landau2}
\end{equation}
Now, since each $q_i$ is a linear combination of the loop momenta, we have the condition
\begin{equation}
\sum_i \alpha_i q_i = 0
\end{equation} with the constraint
\begin{equation}
\alpha_i \geq 0.
\end{equation}
{Eqs.~\ref{eq:landau1} and~\ref{eq:landau2} are the Landau equations~\cite{Landau:1959fi}.
If some of the $\alpha_i = 0$, then the corresponding internal lines get contracted to a vertex. Such a contraction leads to formation of effective vertices which can then be described via an EFT.} Further, in the 
Alpha-parameter space, the end points are 0 and $\infty$. $\alpha_i = 0$ $\forall \ i$ corresponds
to an end-point singularity. The relevant polynomial 
to analyze in this case is the $D$ polynomial in Eq.~\ref{eq:D pol} and its singular points.

Given the Feynman graph of the process, one can define a separation between the vertices in terms of the momentum carried by the connecting lines as  
\begin{equation}
\Delta_i \propto \alpha_i q_i.
\end{equation}
If $\alpha_i \neq 0$, Eq.~\ref{eq:landau1} requires that $q_i^2 = m_i^2$ and we see that an on shell particle propagates from one vertex to the other, that is the reason for divergence of the integrand is if some of the internal lines become onshell. This process of setting some internal lines on-shell, referred to as 
performing unitary cuts, produces a set of sub-integrals called cut integrals. The resulting cut diagrams must 
now describe processes where on-shell particles propagate from one vertex to the 
other~\cite{Sterman:2004pd}. By finding an appropriate subset of these cut diagrams, it is possible to evaluate 
the original integral~\cite{Rozowsky:1997dm}. The parameter $\alpha_i$ is identified with the proper time divided by the mass of the particle. The fact that $\alpha_i \,\textgreater \, 0$ then implies that the particle is propagating forward in time. An immediate corollary of the above for a closed loop is 
\begin{equation}
\sum_i \Delta_i = 0.
\end{equation}

Therefore, it is evident that the Landau equations are statements for finding the singular points of a polynomial derived from the Feynman Integral. The analysis of Norton and Coleman followed by the work of Libby and Sterman \cite{ Libby:1978qf,Libby:1978bx} shows the connection between singularities and threshold processes which are described by EFTs.

\subsubsection{Regions from Feynman graph in the Alpha Parametric form }

A Feynman graph having $l$-loops, $m$-denominators, and $r$-external momenta $(p_1,.....,p_r)$ in $d$-dimension has the form~\cite{LeePomeransky},
\begin{equation}
I(n) = \int \prod_{i=1}^{l} \frac{d^d l_i}{\pi^{\frac{d}{2}}}\prod_{j=1}^{m}\frac{1}{\left( A_{j}^{ik} l_i \cdot l_k+2B_{j}^{ik}l_i \cdot p_k+C_j\right)^{n_j}}
\end{equation} 
where A,B are respectively $l\times l $, $l\times r$ matrices and $C_j$ are constants.

In Alpha-parametric form, $I(n)$ can be written as,
\begin{equation}
I(n) = \frac{\Gamma(|n|-\frac{ld}{2})}{\prod_{j=1}^{m}\Gamma(n_j)}\int \prod_{j=1}^{m}d\alpha_j \alpha_{j}^{n_j-1}\delta \left(1-\sum_{j=1}^{m}\alpha_j\right)\frac{{\cal U}^{|n|-\frac{(l+1)d}{2}}}{{\cal F}^{|n|-\frac{ld}{2}}}
\end{equation}
where ${\cal U}$, ${\cal F}$ are the Symanzik polynomials (of degree $l$ and $(l+1)$ respectively) and $\vert n \vert= n_1 + n_2 + \cdots + n_m$.

In terms of the Symanzik polynomials, the Landau equations can be written as ~\cite{Eden:1966dnq},
\begin{equation}
	{\cal F} = 0,
\end{equation}
\begin{equation}
	\frac{\partial {\cal F}}{\partial \alpha_i} = 0.
\end{equation}
Therefore, we seek approximate solutions, of the type $\alpha_i \sim  c \rho^{v_i}$, of the Landau equations, near the singular surfaces. In our notation $\rho$ is the threshold expansion parameter and $c$ is a 
constant. The sets of $\lbrace v_i \rbrace$ corresponding to each of the solution (at leading order) 
branches, represent all the regions associated with the integral. We will extract these leading order solution 
in the neighborhood of the singular points using the techniques of power geometry, which we 
discuss next. 


\subsubsection{Newton Polytope and Power Geometry}
\label{sect:power_geo}

We are interested in obtaining the leading order scaling of the Alpha-parameters with respect to the
expansion parameter $\rho$. This can be achieved using ideas from power geometry, developed by Bruno
and Bathkin, which allows for obtaining solutions of a polynomial~\cite{bruno2015,bruno2012,bruno:book} in certain limits. 

Consider a generic polynomial in $n$-variables, 
\begin{equation}
\label{def:poly}
g(X)=\sum g_Q X^Q, \quad Q\in S(g)
\end{equation}
where $X=\lbrace x_1,x_2,....x_n\rbrace$ and $Q = \lbrace{Q_1, Q_2, \cdots Q_n \rbrace}$, where $Q_i$ 
are the exponents of the variables $x_i$ for each monomial, i.e., of the form $x_1^{Q_1} x_2^{Q_2} \cdots$
and the $Q_i$s are a set of natural numbers. Let $\chi = \lbrace X^0 \rbrace$ be a set of points such that
$g(X^0) = 0$. If it turns out that $\partial g(\tilde{X}^0) = 0$, where $\tilde{\chi} = \lbrace \tilde{X}^0 \rbrace$ 
and $\tilde{\chi} \D\subseteq \chi$, then the set $\tilde{\chi}$ contain the singular points of the polynomial. 
Power geometry allows one to obtain solutions to polynomials and is particularly useful around singular
points. Before we outline the procedure for obtaining solutions to polynomials using the techniques 
developed by Bruno~\cite{bruno2015,bruno2012,bruno:book}, we briefly summarize the basic definitions and concepts which we have used in our 
subsequent analysis.

\begin{figure}[hbtp]
\centering
\includegraphics[scale=0.9]{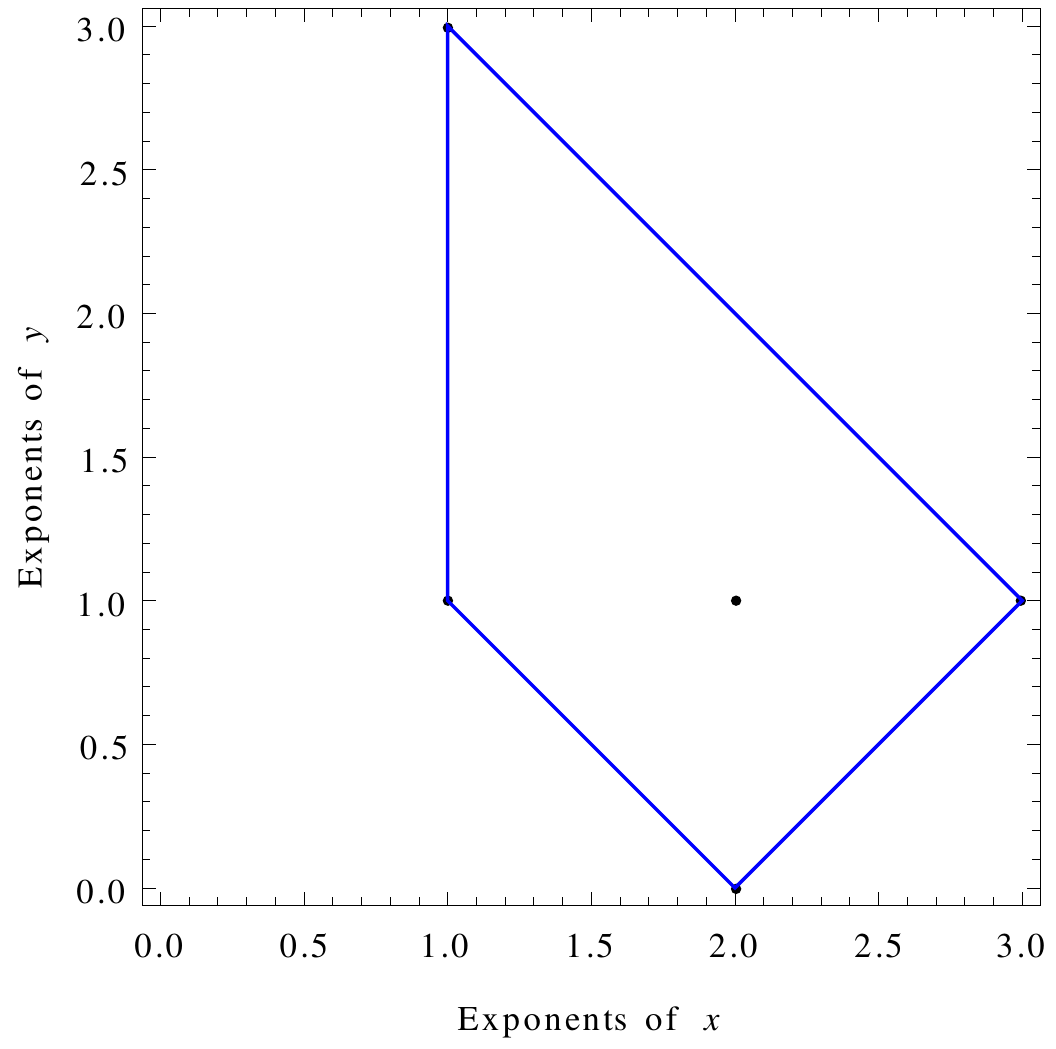}
\caption{Newton polytope of support}
\label{fig:np}
\end{figure}

Let us now define the following:

\begin{enumerate}[label=(\roman{enumi})]
	\item \label{defn1} \emph{Support}: The support $S(g)$ is defined as the set of all vector exponents. For example, given a polynomial in two variables $(x, y)$
	\begin{equation}
		g(x,y) = x y+x^2+x^2 y+ x y^3+x^3 y
		\label{eq:polyex}
	\end{equation}
	$X = \lbrace x,y\rbrace$, and $S(g) = \lbrace (1,1), (1,3), (3,1),(2,0),(2,1) \rbrace$. 

	\item \label{defn2} \emph{Newton Polytope}: The Newton Polytope or Newton Polyhedron is the convex hull of the support $S(g)$. The convex
	hull for the support of the polynomial in Eq.~\ref{eq:polyex} is shown in Fig.~\ref{fig:np}
	
	\item \label{defn3} \emph{Generalized faces}: The \emph{boundary subsets} $\lbrace S^\prime \rbrace$ 
	of the Newton Polytope are its faces $\Gamma_j^d$, where $d$ is the dimension and $j$ labels the face 
	(see Fig.~\ref{fig:np}).

	\item \label{defn4} \emph{Normal Cone}:  Let $S \in \mathbb{R}^n$ be a compact convex set which is 
	the support $S(g)$, with faces $\lbrace S^\prime \rbrace = \lbrace S^\prime_1, S^\prime_2 \cdots S^\prime_r \rbrace$. Let $\ket{\xi}$ be a vector in $\mathbb{R}^n$. Then, $\D \limsup \lbrace \innerproduct{\xi}{\eta}\, \big \vert \,\ket{\eta} \in S \rbrace$ is attained by a vector $\ket{v}$ belonging to a face $S^\prime_i$ for some $i$. It is easy to see that if $\ket{v^\prime}$ is another vector in $\lbrace S^\prime \rbrace$, then $\innerproduct{\xi}{v^\prime} = \innerproduct{\xi}{v}$. The set of all $\ket{\xi}$ such that $\innerproduct{\xi}{v} \ge \innerproduct{\xi}{\eta} \, \text{for}\, \ket{\eta} \in S$ is defined as the normal cone $U_i^d$, where $i$ denotes the $i^{\rm th}$ face and $d$ its dimension.
	 	
	\item \label{defn5} \emph{Cone of the problem}: The \emph{Cone }of the problem is a convex cone of vectors $K = (s_1,..,s_n)$ such 
	that curves of the form
\begin{equation}
x_1 = a_1 t^{s_1} \quad x_2 = a_2 t^{s_2}\quad...\quad x_n = a_n t^{s_n},
\label{convCone}
\end{equation}
where $t$ parametrizes the polynomial, fill those regions of the $X$-variables space that we are interested in.
	\item \label{defn6} \emph{Truncated polynomial}: The truncation of the sum  on the boundary subset is defined as
\begin{equation}
\hat{g}_j^{(d)}=\sum g_Q X^Q \quad Q\in S_j^\prime.
\label{Truncation}
\end{equation}
Such a truncated polynomial should be quasi-homogenous, that is for the polynomial $\hat{g}_j^d(X)$, 
there exists $n$ integers $\lbrace w_1,.....,w_n \rbrace$, called weights of 
the variables, such that the sum $w = w_1 Q_1+.....+w_n Q_n$ is the  same for all nonzero monomials of 
$\hat{g}_j^d(X)$, where $Q = \lbrace Q_1,....,Q_n\rbrace$ is the vector exponent of terms in the polynomial 
$\hat{g}_j^d(X)$. $w$ is then the degree of the polynomial. 
\end{enumerate}

The truncated polynomials corresponding to the faces of the convex hull in Fig.~\ref{fig:np}, as 
well as its weights are listed in Table \ref{Tab:np}.

\begin{table}[h]
\centering
\begin{tabular}{|p{0.7cm}|p{1.9cm}|p{2cm}|p{3cm}|}
\hline
Face & Boundary Subset & Truncated Polynomial & Weights of the polynomial $(w_1, w_2)$\\
\hline
1 & $S_1^{'1}$:(1,1),(1,3) & $x y +x y^3$ &$(a,0)$, where $a$ is an integer\\
\hline
2 & $S_2^{'1}$:(1,3),(3,1) & $x y^3+x^3 y$ & $(a,a)$ \\
\hline
3 & $S_3^{'1}$:(3,1),(2,0) &$x^3 y+x^2$ & $(a,-a)$ \\
\hline
4 & $S_4^{'1}$:(2,0),(1,1) & $x^2+x y$ & $(a,a)$ \\
\hline
\end{tabular}
\caption{A Table showing the Boundary subsets and the associated normals of Newton Polytope}
\label{Tab:np}
\end{table}

Finally, the algorithm to obtain the leading order solution in terms of a parameter $t$, given a polynomial $g(X)$ can be summarized as follows:

\begin{enumerate}
\item The set of singular points (or regular zeros) once identified, can lie on a generalized surface. It turns 
out that it is very
convenient to change variables such that the singular points now lie at the origin, coordinate axis or 
on a coordinate plane.  Such a choice allows one to define the cone of the problem, $K$, easily. As a result 
of these transformations, the polynomial 
$g(X) \rightarrow g^\prime(X^\prime)$, where $X^\prime = T X$ under the map $T$. The problem now 
reduces to analyzing the  $g^\prime(X^\prime)$ polynomial.

\item  The support of $g^\prime(X^\prime)$, denoted as $S(g^\prime(X^\prime))$ is determined and convex hull of the support defines the Newton polyhedron.

\item For each of the two dimensional faces, $\Gamma_j^2$ of the Newton polytope, the normal vectors are
determined, leading to the construction of the normal cone.

\item Only those normal vectors that intersect with the cone of the problem are retained. 

\item For each normal cone lying in the cone of the problem, the truncated polynomial is determined, where
the truncated polynomial is defined on the faces of the convex hull.

\item Finally, the vector $\mathbf{P}$ which gives us the scaling of the variables with respect to the parameter $t$ is obtained, using the following theorem~\cite{bruno2012}: 
{

\begin{theorem}
\label{theorem1}
	If for $t \rightarrow \infty$ the curve
	\begin{equation}
	x = a t^{p_1}\left(1+{\cal O}(1)\right),\, y =  b t^{p_2}\left(1+ {\cal O}(1)\right),\, z = c t^{p_3}\left(1+ {\cal O}(1)\right)
	\end{equation} 
where $a, b,c$ and $p_i$ are constants, belongs to the set  $\mathit{g} =\{X:g(X)=0\}$ and the
vector $P=(p_1,p_2,p_3)$ belongs to $ U_j^d$, then the first approximation $x = a t^{p_1},  y = b t^{p_2}, z = c t^{p_3}$
of the curve satisfies the truncated equation $\hat{g}_j^d(X)=0$.
\end{theorem}}

\end{enumerate}
As a result of the theorem, we get the leading order behavior of the solution. While the method is 
particularly useful for solutions around singular points where the implicit function theorem fails, this method 
can also be applied to obtain solutions about any zero of the polynomial, including regular zeros and has 
the advantage of being easily automated on a computer. Therefore, we will not make the 
distinction between the  
singular and the regular zeros of the polynomial. An important step in this algorithm is to determine the set 
of transformations that map the zeros of the polynomial to either the origin or the coordinate axes or the 
coordinate plane. 
{The Gr\"{o}bner basis ~\cite{bberger} of the Landau 
equations conveniently gives us the required set of transformations as well as the appropriate neighborhoods of singular points where one needs to perform a leading order expansion of the polynomial.} 

With all these key definitions and ideas in place, we go on to enumerate the steps in our 
algorithm, ASPIRE,  that tailors the techniques of power geometry to determining regions in 
section~\ref{sect:derivAlgo} and we discuss the corresponding Mathematica package that automates the 
determination of region.

\subsubsection{Algorithm: ASPIRE}
\label{sect:derivAlgo}

The algorithm proposed by Bruno and Batkhin~\cite{bruno2015,bruno2012,bruno:book} is a very powerful 
method for obtaining the asymptotic behavior of algebraic curves near singular points. For our purposes 
however, we need to extract only the leading order scaling behavior of the alpha parameters with respect
to $\rho$ which is the expansion parameter. Therefore, using  Bruno's theorem~\cite{bruno2012}, we can 
conclude that the truncated polynomials must have a solution of the form where:
\begin{equation}
\alpha_i = a_i \rho^{p_i}(1+{\cal O}(1))
\end{equation}
where $\rho = 1/t$ so that as $t \rightarrow \infty$, $\rho \rightarrow 0$. 
In cases when multiple Alpha parameters scale differently one has to use an approach that systematically finds the leading order expansions of solutions of the Landau equations and thus revealing all the regions. Using the results of Lee and Pomeransky~\cite{LeePomeransky}, we obtain the following parametric form of a Feynman integral \footnote{The choice ${\cal G} = {\cal F} + {\cal U}$ has been used in a recent publication \cite{Semenova:2018cwy} in the context of MoR. In this work, the equivalence of the relevant Newton Polytopes arising in the Feynman parametrization and in the Lee-Pomeransky representation of the Feynman integral has been studied.},

\begin{equation}
\frac{\Gamma(\frac{d}{2})}{\Gamma((l+1)\frac{d}{2}-|n|)\prod_{j=1}^{m}\Gamma(n_j)}\int_0^\infty \cdots \int_0^\infty\prod_{j=1}^{m} d\alpha_j \alpha_{j}^{n_j-1}{\cal G}^{-\frac{d}{2}}
 \label{eq:Gparam}
\end{equation}
where 
\begin{equation}
{\cal G} = {\cal F} + {\cal U} 
\end{equation}

Since $\cal G$ is not quasi-homogeneous in general, one can obtain faces of the
 convex hull of the support of $\cal G$ that correspond to different planes, which in the end  leads to 
 different regions. However, we note that the singular points of ${\cal F}$ now become regular zeros of ${\cal G}$. 

We now enumerate the steps of our algorithm.
\begin{enumerate}
\item Construct the polynomial ${\cal G} = {\cal F} + {\cal U}$.

\item Find the Gr\"{o}bner Basis for the set of Landau equations for ${\cal F}$. 

\item For every neighborhood in the Alpha-parameter space, perform linear transformations to map the 
nearest solution curves of the Gr\"obner Basis elements to the origin, coordinate axis, plane.

\item Using the definition of the small threshold expansion parameter ($x$), in terms of the kinematic invariants, re-express all the constant coefficients like mass and external momenta appearing in the 
above equations.

\item For every transformation applied to the Alpha-parameters, find the support of ${\cal G}$ which has the structure, $S({\cal G})=(Q_0,Q)$, where $Q_0$ is the vector exponent of the small expansion parameter $x$ and $Q=(q_1,q_2,...,q_n)$ are vector exponents of the Alpha-parameters.

\item Find the convex hull of the support.

\item Find the boundary subset for every facet of every Newton polytope.

\item Find the normal cone for each of the facets. This amounts to finding the normal vector to the surfaces.

\item Using the theorem from Bruno~\cite{bruno2012} we conclude that the above truncated polynomials are satisfied by the following expressions for the alpha-parameters

\begin{equation}
\alpha_1 = a_1 x^{p_1}, \alpha_2 = a_2 x^{p_2},...,\alpha_n = a_n x^{p_n}.
\label{eq:alphaSol}
\end{equation} 
Here $a_i \in \mathbb{C}$ and the set of $\mathbf{P} = \lbrace p_i \rbrace$ defines the region. The normal to the surface corresponding to the truncated polynomial gives us $\mathbf{P}$.

\end{enumerate}

The scaling of the threshold parameter with respect to itself, which we call the zeroth component of the 
normal, is by definition $1$. This is ensured by simply 
rescaling the normal which is possible as long as the zeroth component is not zero. 

Jantzen et. al.~\cite{Jantzen:2012mw} attribute the fact that the potential and the Glauber regions were 
missing in ASY,  to the cancellations amongst the Alpha-parameters themselves and thus try to resolve it by 
performing transformations in the Alpha-parameter space which eliminate all differences between the 
Alpha-parameters. In our algorithm we identify these transformations by studying the Gr\"{o}bner basis 
elements. In section~\ref{sect:examples}, we demonstrate the working of our algorithm 
ASPIRE via examples. 

 It is worth noting here that the set of scalings we obtain correspond to asymptotic expansions near the singular points. If the expansion corresponds to regions where the Alpha parameters are far from zero then Norton-Coleman analysis tells us that the Landau equations can be satisfied only if the internal lines are put on-shell which corresponds to a physical particle traveling from one vertex to another. However, if the expansion is in a region where the Alpha-parameters are zero (or close to zero) then the Landau equations can be easily satisfied and such scenarios correspond to shrinking the internal lines and creating effective vertices, which can be interpreted as the emergence of effective field theories. According to our analysis, the scalings obtained from the bottom facets of the Newton Polytope correspond to the latter case. The physical significance of the scalings from the top surface is left to future investigations.

\section{Demonstration of the algorithm and unveiling the associated regions}
\label{sect:examples}

\subsection{One-loop examples} We first consider some one-loop examples already discussed 
in ref.~\cite{ Jantzen:2012mw}. 

\subsubsection{Example 1: Two-point one loop diagram}
\begin{figure}
\centering

\includegraphics[scale=0.9]{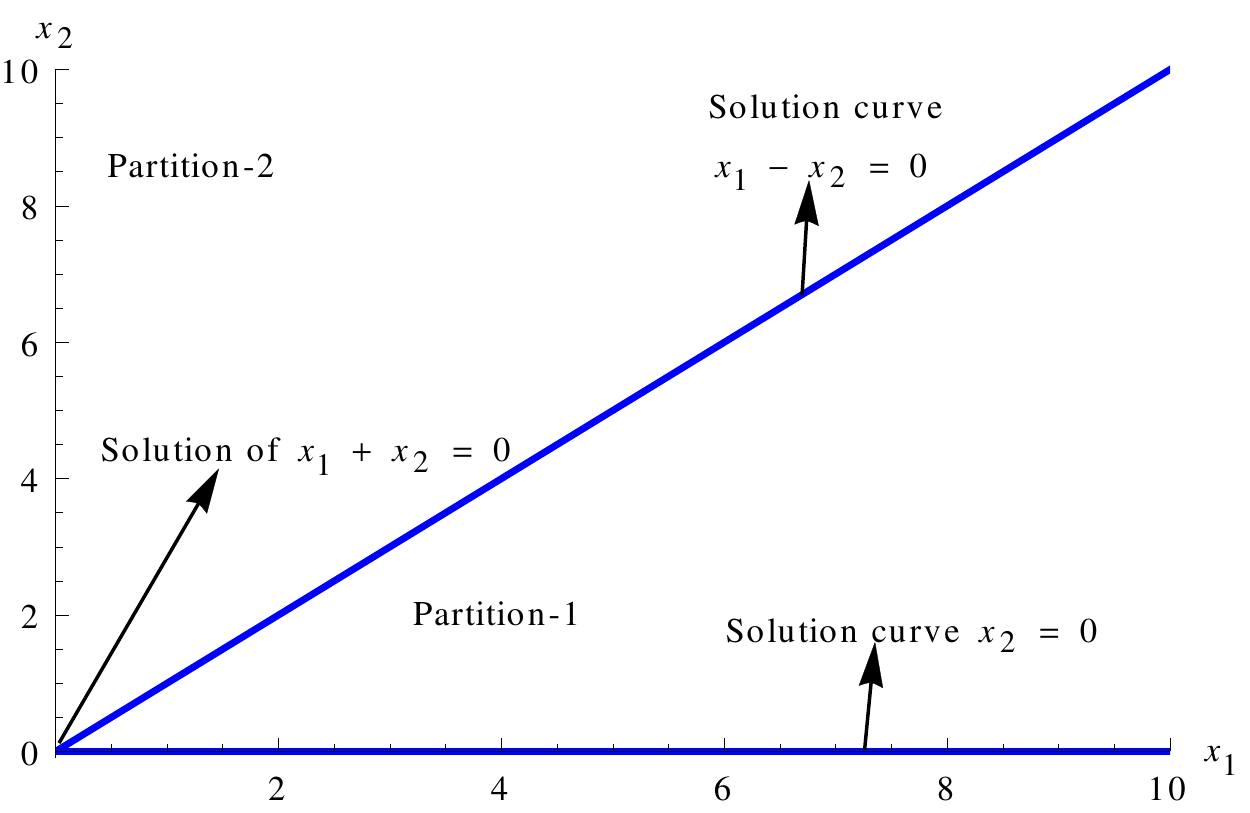}
\caption{Partitioning of alpha parameter space by solution curves of the Gr\"obner Basis elements}
\label{fig:partitioned}
\end{figure}

Consider the integral,

\begin{figure}[H]
\centering
\includegraphics[scale=0.5]{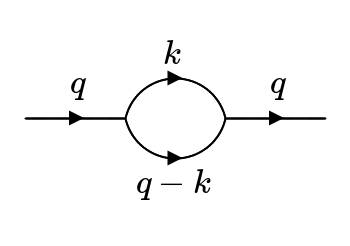}
\caption{A two point one loop diagram}
\end{figure}

\begin{equation}
I=\int \frac{d^d k}{(k^2-m^2)((k-q)^2-m^2)}.
\label{eq:example1}
\end{equation}
Here $q$ is the external momentum, $k$, the loop momentum and the threshold expansion parameter is 
defined as 
$\D y = m^2-\frac{q^2}{4}$. The energy scales involved in this integral are set by: $q$, $\D\frac{y}{q}$ and 
$\sqrt{y}$. From~\cite{beneke_smirnov_1997,Jantzen:2012mw},  the contributing regions are the hard and 
the potential regions whose scaling in the momentum space with respect to the threshold parameter are 
given as, 
\begin{equation}
\textit{Hard region:} \ \left( k_0\sim q ,k \sim q \right)
\end{equation} 
\begin{equation}
\textit{Potential region:} \ \left( k_0 \sim \frac{y}{q},k \sim \sqrt{y}\right), \left(k_0 \sim \sqrt{y}, k \sim \frac{y}{q}\right)
\end{equation}
We will reproduce all those contributing regions using the algorithm developed in section~\ref{sect:derivAlgo}.
We write the integral in Eq.~\ref{eq:example1} in the Alpha-parameter space which gives the ${\cal U}$ and 
${\cal F}$ polynomials using the package \textbf{UF.m}~\cite{UF} (described in the appendix). In the
Mathematica code, this function is called as follows:
\begin{equation}
\text{UF}\left[\{k\},\,\left\{-(k^2-m^2),-((k-q)^2-m^2)\right\},\,\left\{q^2\to \text{qq},m^2\to \frac{\text{qq}}{4}+y\right\}\right] \footnote{One assigns a negative sign to each of the propagators in order to get the correct ${\cal U}$ \cite{Jantzen:2012mw}}
\end{equation} 
yielding an output
\begin{equation}
\left\{x_1+x_2,\,\frac{1}{4} \text{qq}\, x_1^2-\frac{1}{2} \text{qq}\, x_1 x_2+\frac{1}{4} \text{qq}\, x_2^2+x_1^2 y + 2 x_1\, x_2\, y+x_2^2\, y,\,1\right\}
\label{eq:UF_example1}
\end{equation}
The first and second elements are the ${\cal U}$ and ${\cal F}$ polynomials respectively, while the third 
element of the output is the number of loops, which is $1$ in this example. There are two Alpha
parameters, denoted as $x_1$ and $x_2$ corresponding to the two denominators in Eq.~\ref{eq:example1}.
For ease of reference, we list ${\cal U}$ and ${\cal F}$ polynomials below:
\begin{equation}
{\cal U} = x_1+x_2,
\label{eq:example1_u}
\end{equation}

\begin{equation}
{\cal F} = \frac{1}{4} q^2 x_1^2-\frac{1}{2} q^2 x_1 x_2+\frac{1}{4} q^2 x_2^2+y x_1^2+2 y x_1 x_2+y x_2^2
\label{eq:example1_f}
\end{equation}
We wish to find the locations of the singularities in the Alpha-parameter space with the help of Landau equations,
\begin{equation}
{\cal F} = 0,
\end{equation}
\begin{equation}
\frac{\partial{\cal F}}{\partial x_1} = \frac{\partial{\cal F}}{\partial x_2} = 0. 
\end{equation}
Next we find the Gr{\"o}bner basis of the set of Landau Equations for which, we use the Mathematica function GroebnerBasis via the command 
\begin{equation}
\text{GroebnerBasis}\left[\left\lbrace {\cal F},\frac{\partial{\cal F}}{\partial x_1}, \frac{\partial{\cal F}}{\partial x_2} \right\rbrace,\left\lbrace x_1,x_2 \right\rbrace\right]
\end{equation} 
which gives the elements,
\begin{equation}
\mathbb{G} = \lbrace q^2 y x_2,\, (x_1+x_2)y,\, q^2 (x_1-x_2) \rbrace.
\label{eq:gb} 
\end{equation}
The ${\cal F}$ polynomial given in Eq.~\ref{eq:example1_f} can be written in terms of the elements in 
Eq.~\ref{eq:gb}. The simultaneous zeros of ${\cal F}$ and its first order derivatives define the
singular points which in general coincide with the zeros of the Gr\"obner basis
elements. As seen in fig.~\ref{fig:partitioned}, the solution curves of the Gr\"obner basis elements partition the Alpha-parameter space and so one can now choose different neighborhoods for studying the leading behavior of $\mathcal{F}$ or equivalently $\mathcal{G}$. To perform the expansion in the neighborhood of the solution curve of third element of Gr\"obner Basis i.e. $q^2(x_1-x_2)$, we define a set of linear transformations $\left\lbrace x_1 \rightarrow \mathbf{a} x_1^\prime,x_2 \rightarrow x_2^\prime + \mathbf{a} x_1^\prime \right\rbrace $ and 
$\left\lbrace x_1 \rightarrow x_1^\prime + \mathbf{a} x_2^\prime, x_2 \rightarrow \mathbf{a} x_2^\prime \right\rbrace $ 
respectively. Under these transformations $x_1-x_2 \rightarrow x_2^\prime = 0$ and $x_1-x_2 \rightarrow x_1^\prime = 0$ respectively. We can now expand 
${\cal F}$ or equivalently ${\cal G}$ in the variable $x_2^\prime$ or $x_1^\prime$. In all of the above calculations we need to 
keep in mind the constraint $x_i \geq 0$.  These transformations are analogous to the approach in ASY2, where linear 
transformations were performed when the Alpha-parameters appeared with a negative sign between 
them. Such transformations reveal the potential and the Glauber regions.

We now list all distinct transformations:
\begin{itemize}
\item Identity transformation:
\begin{equation}
T_1 \equiv \left\lbrace x_1 \rightarrow x_1, x_2 \rightarrow x_2 \right\rbrace
\end{equation}
\item Non-trivial transformations:
\begin{eqnarray}
T_2 \equiv \left\lbrace x_1 \rightarrow \frac{x_1}{2},x_2 \rightarrow x_2 + \frac{x_1}{2}\right\rbrace\\
T_3 \equiv \left\lbrace x_1 \rightarrow x_1 + \frac{x_2}{2},x_2 \rightarrow \frac{x_2}{2}\right\rbrace
\end{eqnarray}
\end{itemize}

In the above list of transformations, we have fixed the constant $\mathbf{a}$ by ensuring that the transformations leave the delta function in the integral, unchanged. Now we go on to compute ${\cal G}$ with all the above transformations 
$T=\left\lbrace T_1,T_2,T_3 \right\rbrace$, where the ${\cal G} = \lbrace {\cal G}_1, {\cal G}_2, {\cal G}_3 \rbrace$ corresponding to the three transformations. Therefore:
\begin{equation}
{\cal G}_1  \equiv \frac{1}{4} q^2\, x_1^2 - \frac{1}{2} q^2 \, x_1\, x_2 + \frac{1}{4} q^2\, x_2^2 + x \, x_1^2 + 2\, x\, x_1\, x_2 + x_1 + x\, x_2^2 + x_2,
\label{eq:example1_G1}
\end{equation}

\begin{equation}
{\cal G}_2 \equiv \frac{1}{4} q^2\, x_1^2  +x\, x_1^2 + 2\, x\, x_1\, x_2 + x_1 + x\, x_2^2 + x_2,
\label{eq:example1_G2}
\end{equation}
and
\begin{equation}
{\cal G}_3 \equiv \frac{1}{4} q^2\, x_2^2 + x\, x_1^2 + 2\, x\, x_1\, x_2 + x_1 + x\, x_2^2 + x_2 
\label{eq:example1_G3}
\end{equation}
Here, we have substituted $y \rightarrow x$ and $q^2 \rightarrow x^0 $ (i.e. $q^2 \rightarrow 1 $).
We next find the support of the ${\cal G}$ polynomials. Here we consider the threshold parameter, $x$, as 
an independent co-ordinate, and therefore, while extracting the vector exponents of the Alpha-parameters 
$x_i$, we extract the exponents of $x$ as well. 

The support $S_i$ of ${\cal G}_i$, where $i$ enumerates the three polynomials coming from the three
transformations are,

\begin{equation} 
S_1= \left( \begin{array}{ccc}
 0 & 1 & 0 \\
 0 & 2 & 0 \\
 1 & 2 & 0 \\
 0 & 0 & 1 \\
 0 & 1 & 1 \\
 1 & 1 & 1 \\
 0 & 0 & 2 \\
 1 & 0 & 2 
\end{array} \right), \, \,
S_2=\left( \begin{array}{ccc}
 0 & 1 & 0 \\
 0 & 2 & 0 \\
 1 & 2 & 0 \\
 0 & 0 & 1 \\
 1 & 1 & 1 \\
 1 & 0 & 2 
\end{array} \right),\, \
S_3=\left( \begin{array}{ccc}
 0 & 1 & 0 \\
 1 & 2 & 0 \\
 0 & 0 & 1 \\
 1 & 1 & 1 \\
 0 & 0 & 2 \\
 1 & 0 & 2 
\end{array} \right) 
\label{eq:example1_support}
\end{equation}
Each row is a point in $\mathbb{R}^3$ and we denote each row as $P_i$.  The next step in the algorithm
is to determine the convex hull of the support in $\mathbb{R}^3$ for which, we use the function \textbf{CHNQuickHull}~\cite{Petrich} as follows
\begin{equation}
\text{CHNQuickHull}[S].
\end{equation}

\begin{figure}[t]
\begin{center}
\begin{subfigure}{0.3\textwidth}
\includegraphics[width=0.7\linewidth, height=5cm]{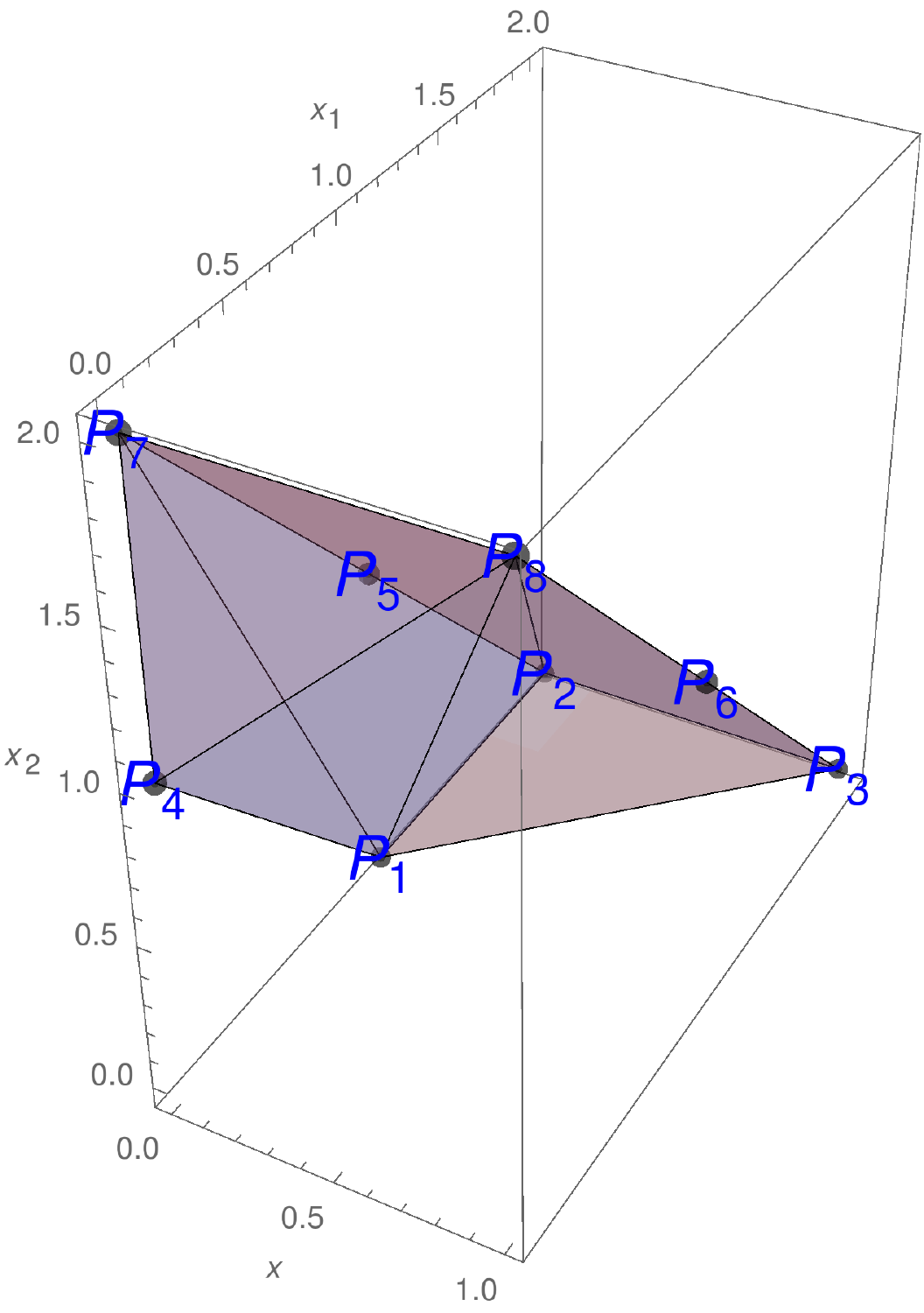} 
\caption{}
\label{fig:np1}
\end{subfigure}
\begin{subfigure}{0.3\textwidth}
\includegraphics[width=0.7\linewidth, height=5cm]{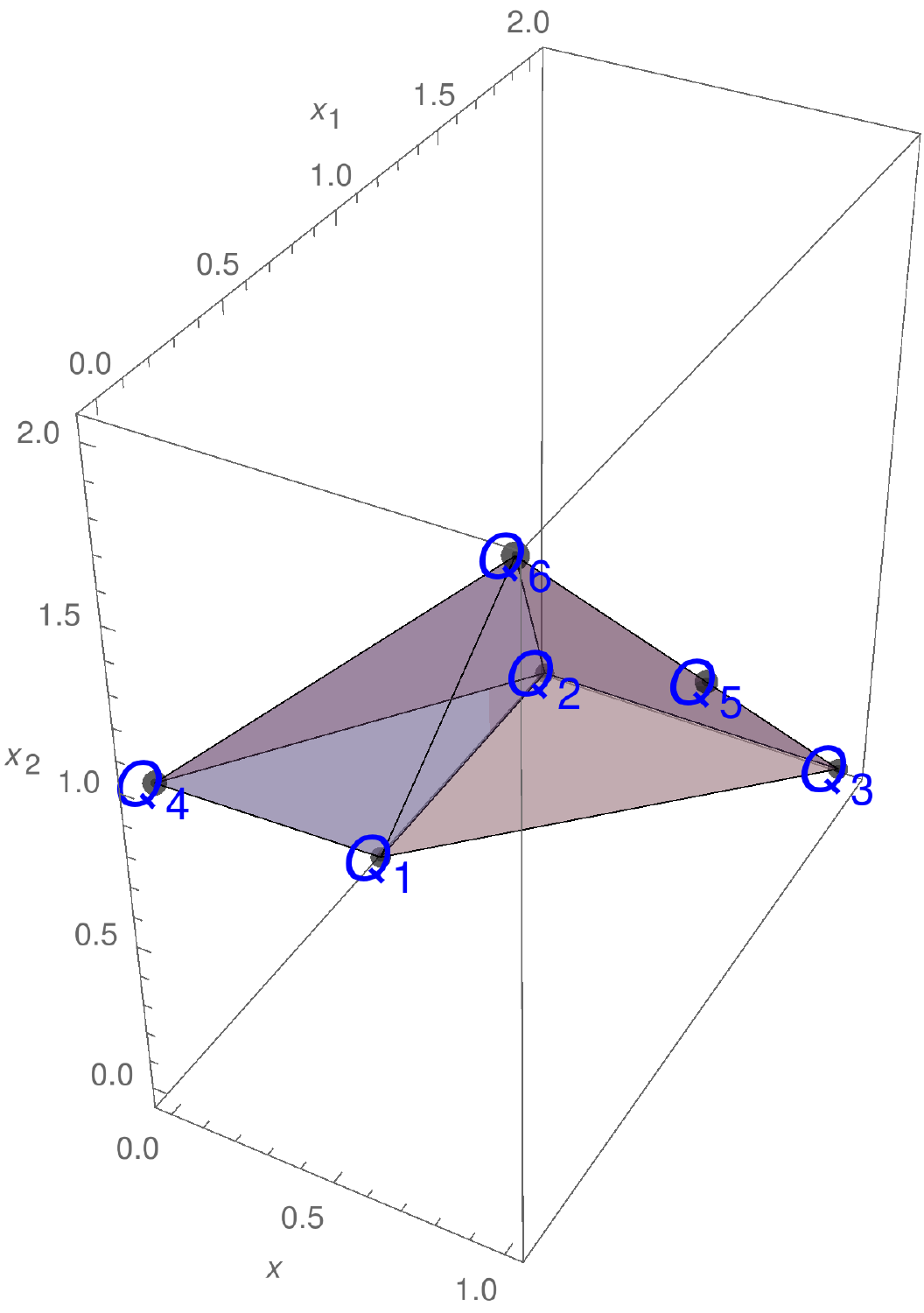}
\caption{}
\label{fig:np2}
\end{subfigure}
\begin{subfigure}{0.3\textwidth}
\includegraphics[width=0.7\linewidth, height=5cm]{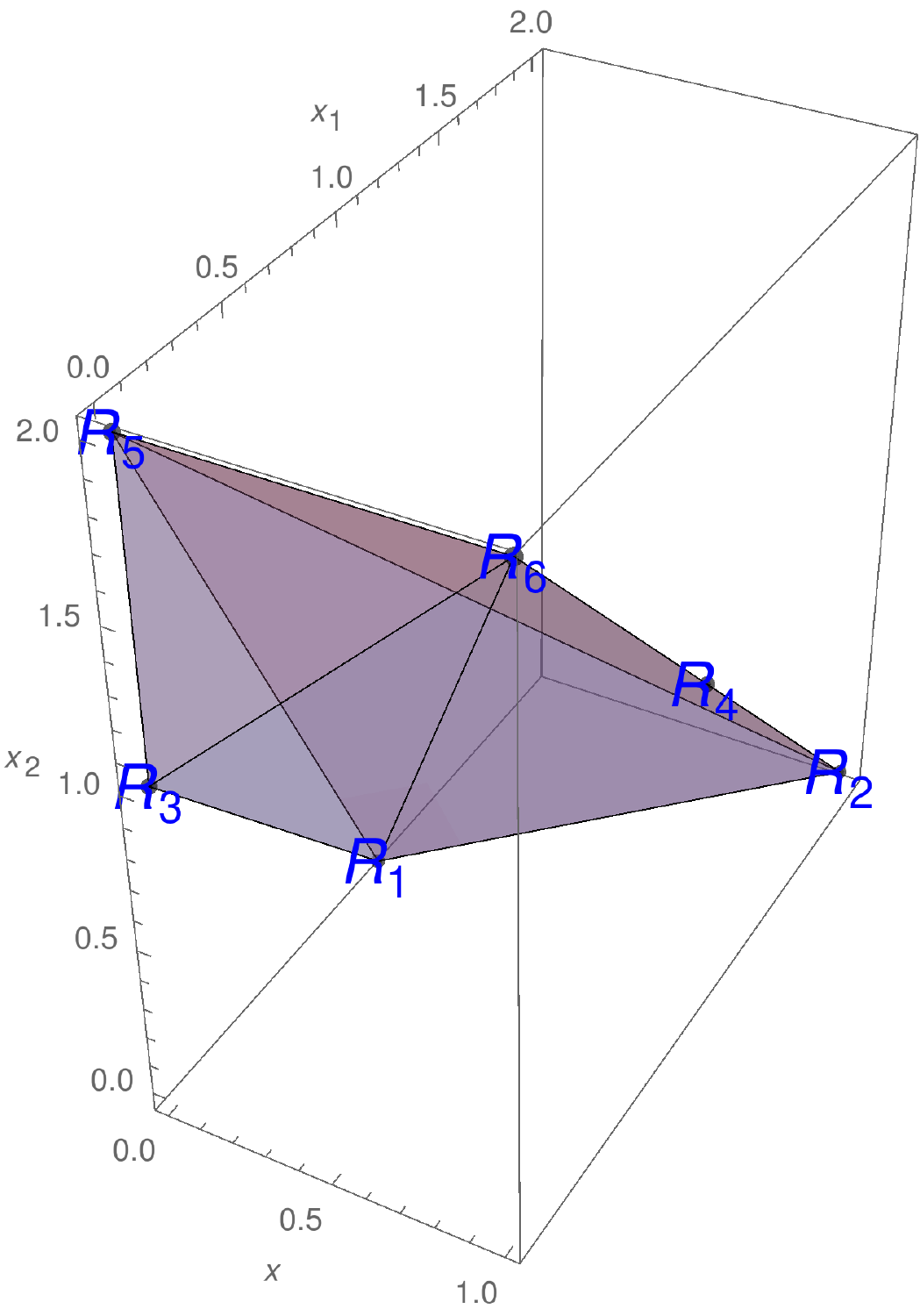} 
\caption{}
\label{np3}
\end{subfigure}
\end{center}
 
\caption{From left to right- Newton Polytopes for ${\cal G}_1$, ${\cal G}_2$ and ${\cal G}_3$.}
\label{fig:NewtonPolytopes}
\end{figure}

Finally, we can obtain the scaling of the Alpha parameters and hence the regions. For ${\cal G}_1$, 
the vertices are,
\begin{equation}
S_1 \equiv \lbrace P_1(0,1,0),P_2(0,2,0),P_3(1,2,0),P_4(0,0,1),P_5(0,1,1),P_6(1,1,1),P_7(0,0,2),P_8(1,0,2)\rbrace
\end{equation}
as seen in Fig.~\ref{fig:NewtonPolytopes} in the left-most panel labelled (a).

We then go on to find the normal vectors corresponding to each of the surfaces using the function 
\textbf{genNormalCoordinates} and obtain
\[\{ \{v(1)\to 0,v(2)\to 0,c\to 0,\text{surf}\to -1\}, \text{Null}, \{v(1)\to -1,v(2)\to -1,c\to -1,\text{surf}\to 1\}\] 
\[ \{v(1)\to 0,v(2)\to 0,c\to 0,\text{surf}\to -1\},\{v(1)\to -1,v(2)\to -1,c\to -1,\text{surf}\to 1\},\text{Null},\text{Null},\text{Null}\}\]
where $v(1), v(2)$ are the components of the normal vector of the facets of the Newton Polytope and $c$ is 
a constant. The presence of the element \emph{Null} implies that the code was not able to find any normal 
vector based on the conditions \ref{top} and \ref{bottom}. 
In our code, we determine the normal vector corresponding to the facets of the Newton polytope based on following conditions :
\begin{enumerate}
\item $\vec{r}.\vec{v}= c$ and $ \vec{r'}.\vec{v} < c $, where $\vec{r}$ belongs to a boundary subset of Newton polytope and $\vec{r'} $ does not. We name the surface giving the normal vector depending on this condition ``the top facet" and assign a label ``$surf\rightarrow 1$" for that surface. \label{top}
\item $\vec{r}.\vec{v}= c$ and $ \vec{r'}.\vec{v} > c $. For this condition, we call the surface giving the normal vector ``the bottom facet" and label the surface by ``$surf\rightarrow -1$". \label{bottom} 
\end{enumerate}

We use the function \textbf{UniqueRegions} (explained in the Appendix) to select only the unique 
normal vectors and hence the unique regions. In the above list, we only have one region, which is the hard 
region $\lbrace 0,0 \rbrace$.

For the polynomial ${\cal G}_2$, we have six vertices,
 \begin{equation}
 S_2 \equiv \lbrace Q_1(0,1,0),Q_2(0,2,0),Q_3(1,2,0),Q_4(0,0,1),\\ Q_5(1,1,1),Q_6(1,0,2)\rbrace 
 \end{equation}
and the corresponding convex hull is seen in the center panel labelled (c) in Fig.~\ref{fig:NewtonPolytopes}.
The normal vectors for the different faces of the hull are,
\[\Bigg\{\{v(1)\to 0,v(2)\to 0,c\to 0,\text{surf}\to -1\},\text{Null},\{v(1)\to -1,v(2)\to -1,c\to -1,\text{surf}\to 1\},\] \[ \{v(1)\to -1,v(2)\to -1,c\to -1,\text{surf}\to 1\},\text{Null},\left\{v(1)\to -\frac{1}{2},v(2)\to -1,c\to -1,\text{surf}\to -1\right\}\Bigg\}\]
out of which we get two unique normal vectors \[\{v(1)\to 0,v(2)\to 0,c\to 0\}\] and
\[\left\{v(1)\to -\frac{1}{2},v(2)\to -1,c\to -1\right\}\] from the bottom facets.  Therefore, we have two
regions are $\lbrace0,0\rbrace$ and $\lbrace-\frac{1}{2},-1\rbrace$.

Similarly, for the support of ${\cal G}_3$
\begin{equation}
S_3 \equiv \lbrace R_1(0,1,0),R_2(1,2,0),R_3(0,0,1),R_4(1,1,1,),R_5(0,0,2),R_6(1,0,2) \rbrace 
\end{equation}
the convex hull is seen in the right-most panel labelled (c) in Fig.~\ref{fig:NewtonPolytopes} and the unique 
regions obtained from the normal vectors to the facets are, $\lbrace0,0\rbrace$ and $\lbrace -1,-\frac{1}{2}\rbrace$, where both are obtained from the bottom facets.

Finally, taking the union of all of the above regions we have the following set of regions:
\begin{equation}
\left(
\begin{array}{c|c}
Normal & Facet: top/bottom(1/-1) \\
\hline

 \{0,0\} & \{-1\} \\
 \{-1,-1\} & \{1\} \\
 \left\{-\frac{1}{2},-1\right\} & \{-1\} \\
 \left\{-1,-\frac{1}{2}\right\} & \{-1\} \\
\end{array}
\right)
\end{equation}
where the first entry with $(0,0)$ scaling corresponds to the hard region, while the other two $\left\{-\frac{1}{2},-1\right\}$ and $\left\{-1,-\frac{1}{2}\right\}$ are the
potential regions, in agreement with ASY/ASY2. As mentioned earlier, we defer the discussion of the
components of the normal vectors obtained from the top region to a future publication.
 
\subsubsection{Example 2: A five point one loop diagram}
\begin{figure}[H]
\centering
\includegraphics[scale=0.25]{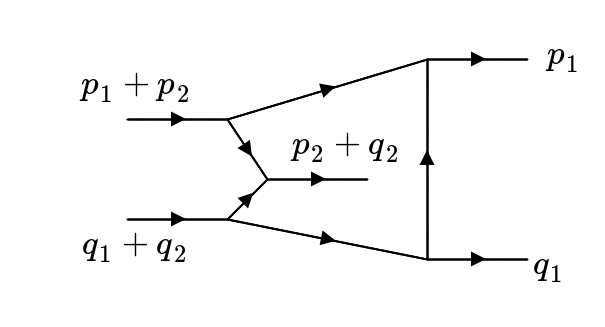}
\caption{A five point one-loop diagram 
}
\label{fig:example2_5pt}
\end{figure}

In this example, we consider the following one loop five point integral seen in Fig.~\ref{fig:example2_5pt}, 
which has also been discussed in~\cite{Jantzen:2012mw}.
\begin{equation}
I(Q^2,m^2;d)=\int d^dk\frac{1}{(k^2-m^2)(k^2-2pk)(k^2+2pk)(k^2-2qk)(k^2+2qk)}
\end{equation}
Here we have $p_1=p_2=p$ and $q_1=q_2=q$. At threshold, $p^2 \rightarrow 0$, $q^2 \rightarrow 0$, 
$2\,p\,q \rightarrow Q^2$ where $Q^2$ is the hard scale and $m^2 \ll Q^2$. The threshold expansion parameter is, $x=m^2/Q^2$.

The Symanzik polynomials are,
\begin{eqnarray}
{\cal U} = x_1+x_2+x_3+x_4+x_5 \qquad \qquad \qquad \qquad \qquad \qquad \qquad \quad\\
{\cal F} = m^2x_1^2+m^2x_1x_2+m^2x_1x_3+m^2x_1x_4+m^2x_1x_5+Q^2x_2x_4-Q^2x_3x_4-Q^2x_2x_5+Q^2x_3x_5
\end{eqnarray}
and the Gr{\"o}bner basis elements for the Landau equations are given by,
\begin{equation}
\mathbb{G}=\lbrace Q^2(x_4-x_5), m^2Q^2(x_3+x_5),Q^2(x_2-x_3), m^2(x_2+x_3+x_4+x_5),m^2x_1 \rbrace.
\end{equation}
From the elements of the Gr\"obner basis we can immediately conclude that we will need transformations 
listed in Table~\ref{Table:5p1l}

\begin{table}[H]
\centering
\begin{tabular}{|c|c|}
\hline
Element & Transformation \\
\hline
\multirow{2}{*}{$Q^2(x_4-x_5)$} & $(x_4 \rightarrow x_4+ax_5)$, $(x_5 \rightarrow ax_5)$\\
\hhline{~-}
& $(x_5 \rightarrow x_5+ax_4)$, $(x_4 \rightarrow ax_4)$\\
\hline
$m^2Q^2(x_3+x_5)$ & None required\\
\hline
\multirow{2}{*}{$Q^2(x_2-x_3)$} & $(x_2 \rightarrow x_2+ax_3)$, $(x_3 \rightarrow ax_3)$\\
\hhline{~-}
& $(x_2 \rightarrow x_2+ax_3)$, $(x_3 \rightarrow ax_3)$\\
\hline 
$m^2(x_2+x_3+x_4+x_5)$ & None required\\
\hline
$m^2x_1$ & None required\\
\hline
\end{tabular}
\caption{Mapping Gr\"obner Basis elements to coordinate origin, plane or curve via linear transformations}
\label{Table:5p1l}
\end{table}

Next we apply these transformations to the ${\cal G}$ polynomial that we construct from the Symanzik polynomials to get four versions for each of the four transformations. We find the Newton Polytope of the support for each of the ${\cal G}$ polynomials and in each case, we determine the normal vectors listed 
as follows:
\begin{equation}\left(
\begin{array}{c|c}
Normal & Facet: top/bottom(1/-1) \\
\hline
 \{-1,-1,-1,-1,-1\} & \{1\} \\
 \{-1,-1,-1,0,0\} & \{-1\} \\
 \{-1,0,0,-1,-1\} & \{-1\} \\
 \{0,0,0,0,0\} & \{-1\} \\
  \{-1,0,0,0,0\} & \{1\} \\
 \{-1,0,-1,-1,-1\} & \{-1\} \\
 \{-1,-1,-1,0,-1\} & \{-1\} \\
 \{-1,-1,-1,-1,0\} & \{-1\} \\
  \{-1,-1,0,-1,-1\} & \{-1\} \\
\end{array}
\right)
\end{equation}

Once we have the regions in the Alpha-parameter space, we identify the leading order expansions of the Symanzik Polynomials with the corresponding expansion of propagators in the momentum space. 
We recover the following regions corresponding to the bottom facet:
\begin{enumerate}[label=\roman*)]
\item The hard region : $\lbrace 0,0,0,0,0 \rbrace$
\item The collinear regions : $ \lbrace -1, -1, -1, 0,0 \rbrace$ and $\lbrace -1, 0,0,-1,-1\rbrace$ 
\item The Glauber regions: $\lbrace -1,-1,-1,0,-1\rbrace$ and $\lbrace -1,0,-1,-1,-1\rbrace$ 
\item The Scaleful regions: $\lbrace -1,-1,-1,-1,0 \rbrace$ and $\lbrace -1,-1,0,-1,-1 \rbrace$
\end{enumerate}

The scalings $\lbrace -1,-1,-1,-1,-1\rbrace$ and $\lbrace-1,0,0,0,0\rbrace$ from the top facets do not correspond any physical region.

The evaluation of the contribution from Glauber regions  $\lbrace -1,-1,-1,0,-1\rbrace$ and $\lbrace -1,0,-1,-1,-1\rbrace$ need additional analytic regularization as is often the case in case in SCET and has been discussed in refs.\cite{Jantzen:2012mw,Becher:2011dz}.

It needs to be noted here that the regions we obtained above correspond to choosing a specific orientation of the Newton Polytope (for a discussion on rotation of the Newton Polytope in the alpha parametric space the reader is referred to ~\ref{sect:rotation}). It is possible, by arbitrary rotations, to generate an infinite set of scaleless/scaleful regions. However, for a given orientation there is only a finite set of regions. In our implementation, we determine the unique set of regions for a fixed orientation.

\subsection{A two-loop fish diagram}

\begin{figure}[H]
\centering
\includegraphics[scale=0.25]{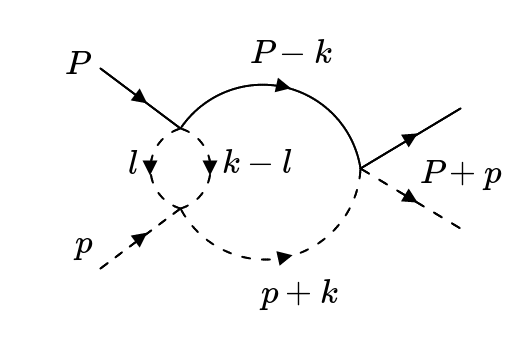}
\caption{A Two loop fish diagram in $\pi$-K scattering}
\label{fig:2Lfish}
\end{figure}

We apply our technique for the two-loop fish diagram, seen in Fig.~\ref{fig:2Lfish}, that has been discussed 
in~\cite{KaiserSchweizer} in the context of expansion by regions for the $\pi -K$ scattering at the threshold. This process can be studied as the scattering of a pion having mass $m$ and a kaon having mass $M$. The momenta of the pion and kaon are $p$ and $P$ respectively.

For this type of diagram, there are two mass scales ($m$ and $M$) based on which the scalings of two loop 
momenta, $k$ and $l$, one can have the following five possible regions as discussed: $\left(k\sim M,l\sim M,k-l\sim M \right)$, 
$\left(k\sim M,l\sim M,k-l\sim m\right)$, $\left(k\sim M,l\sim m \right)$, $\left(k\sim m,l\sim M\right)$, 
$\left(k\sim m,l\sim m \right)$.

In~\cite{KaiserSchweizer}, it has been discussed that the h-h region starts contributing at order one while the s-s region at order $\D\frac{m}{M}$. The h-h' and h-s regions contribute at order $\D\frac{m^2}{M^2}$.

The integral having an internal pion loop is given by
\begin{equation}
	I = \int \frac{d^dk d^dl}{(m^2-l^2)(m^2-(k-l)^2)(2 P k-k^2)(-2 p k-k^2)}
\label{eq:2Lfish}
\end{equation}

At threshold, $p^2 \rightarrow m^2$, $P^2\rightarrow M^2$ and 
$(p+P)^2=(m+M)^2$. The expansion parameter is $x=\frac{m}{M}$.
 
For the integral in 
Eq.~\ref{eq:2Lfish}, we get the following Symanzik polynomials,
\begin{equation}
\begin{aligned}
{\cal U} ={} & x_1 x_2+x_1 x_3+x_2 x_3+x_1 x_4+x_2 x_4  \\
\end{aligned}
\end{equation}
\begin{equation}
\begin{aligned}
{\cal F} ={} & x_1 x_2^2 x^2+x_1 x_4^2 x^2+x_2 x_4^2 x^2+x_1^2 x_2 x^2+x_1^2 x_3 x^2+x_2^2 x_3 x^2+ \\
             & 2 x_1 x_2 x_3 x^2+x_1^2 x_4 x^2+x_2^2 x_4 x^2+2 x_1 x_2 x_4 x^2- 2 x_1 x_3 x_4 x-2 x_2 x_3 x_4 x+x_1 x_3^2+x_2 x_3^2 
\end{aligned}
\end{equation}

As mentioned in our algorithm, we find the Gr{\"o}bner basis elements of ${\cal F}$ and its derivatives with respect to $x_1,x_2,x_3,x_4$ :
\begin{gather}
\nonumber
\mathbb{G} = \Big\lbrace -(x_3+x (x_2-x_4)) (x_3-x x_4) (x_3-x (x_2+x_4)), x_2 ( x_3+ x (x_2-x_4)) (x (x_2 + x_4)  - x_3),\\
\nonumber
 x_3^2 + x (-x x_2^2 + x x_4^2 + 2(x (x+1) (x_1+x_2)) -x_3 x_4), (x + 1) (x_1 + x_2) (x_3 - x x_4), (x_1+x_2) (x_2 x^2-x_4 x+x_3),\\
\nonumber
(x_1+x_2) (x_3-x x_4) (x_2+x_3+(x+2) x_4), x^2 (x_1^2-x_2^2),(x_1^2-x_2^2) (x_3-x x_4) \Big\rbrace
\end{gather}
We use the elements of the Gr{\"o}bner basis, together with the constraint $x_i \geq 0$, to obtain five 
transformations out of which one is trivial and others are non-trivial and are listed below:
\begin{itemize}
\item Identity transformation :
\begin{eqnarray}
T_1\equiv\lbrace x_1 \rightarrow x_1,x_2 \rightarrow x_2,x_3 \rightarrow x_3,x_4 \rightarrow x_4 \rbrace 
\end{eqnarray}
 
\item Non-trivial transformations :
\begin{eqnarray}
T_2\equiv \lbrace x_1 \rightarrow x_1 + \frac{x_2}{2},x_2 \rightarrow \frac{x_2}{2},x_3 \rightarrow x_3,x_4 \rightarrow x_4\rbrace \\
T_3\equiv \lbrace x_1 \rightarrow \frac{x_1}{2} ,x_2 \rightarrow x_2 + \frac{x_1}{2},x_3 \rightarrow x_3,x_4 \rightarrow x_4\rbrace \\
T_4\equiv \lbrace x_1 \rightarrow x_1 ,x_2 \rightarrow x_2+\frac{x_4}{2},x_3 \rightarrow x_3,x_4 \rightarrow \frac{x_4}{2}\rbrace \\
T_5\equiv \lbrace x_1 \rightarrow x_1 ,x_2 \rightarrow \frac{x_2}{2},x_3 \rightarrow x_3,x_4 \rightarrow x_4 + \frac{x_2}{2}\rbrace
\end{eqnarray}
\end{itemize}
As before, we compute ${\cal G}$ polynomials by applying all of these transformations, determine
the support in each case and the corresponding normal vectors. This leads us to the following list of 
unique normal vectors:
\begin{equation}
\left(
\begin{array}{c|c}
Normal & Facet: top/bottom(1/-1) \\
\hline
 \{-2,-2,-1,-2\} & \{-1\} \\
 \{-2,-2,-2,-2\} & \{1\} \\
 \{-2,0,0,0\} & \{-1\} \\
 \{0,-2,0,0\} & \{-1\} \\
 \{0,0,0,0\} & \{-1\} \\
 \{-2,-2,0,-1\} & \{1\} \\
 \{-2,-2,0,-2\} & \{1\} \\
\end{array}
\right)
\end{equation}

The regions are (once again from the bottom facet) are:
\begin{enumerate}[label=\roman*)]
\item The hard-hard region : $\lbrace 0,0,0,0 \rbrace$
\item The hard-soft region : $ \lbrace -2,0,0,0 \rbrace$ 
\item The soft-soft region: $\lbrace -2,-2,-1,-2 \rbrace$
\item The Scaleful regions : $\lbrace 0,-2,0,0 \rbrace$
\end{enumerate}

The region $\lbrace 0,-2,0,0 \rbrace$ is also isolated by ASY. This region has not been identified in the existing literature to the best of our knowledge, hence, 
we give them a generic name of ``scaleful region".


\subsection{One loop Scalar Triangle Diagrams in Sudakov Limits}
\begin{figure}[H]
\centering\includegraphics[scale=0.25]{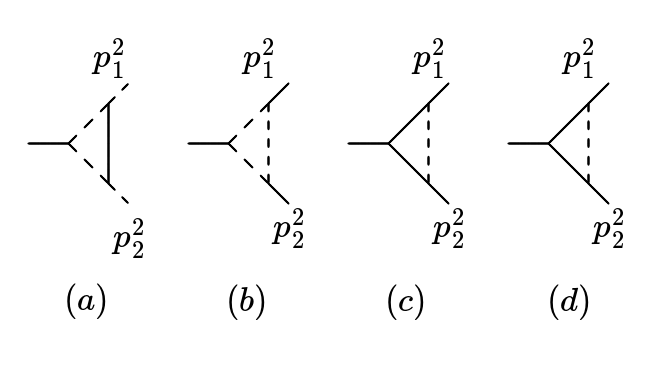}
\caption{One loop Sudakov integrals in limits (a)-(d). The solid and dashed lines represent respectively the massive and the mass-less particles.}
\label{fig:sudakov}
\end{figure}

As a final illustration of our algorithm, we discuss the case of one-loop Sudakov integrals in the following 
limits, illustrated in Fig.~\ref{fig:sudakov}. 
Sudakov limits appear in processes where the square of the momentum
transfer is large compared to the squares of the masses.
A detailed discussion may be found in the chapter on Sudakov
limits in ref.~\cite{Smirnov_applied}.  See also ref.~\cite{Softbook}.

\begin{enumerate}[label=(\alph*)]
\item \label{lim_a} On-shell massless fermions ($p_1^2=p_2^2=0$) and gauge bosons with small non-zero mass, $m^2 \ll -s \equiv Q^2$. We also choose,
\begin{equation}
\label{def:mom}
p_{1,2} = (Q/2,0,0,\mp Q/2)
\end{equation} 

so that $2\, p_1 \cdot p_2 = Q^2$.

\item Massless gauge bosons and off-shell massless fermions ($p_1^2=p_2^2=-M^2), M^2 \ll -s$. We use,
\begin{equation}
p_{1,2}=\tilde{p}_{1,2}-\frac{M^2}{Q^2}\tilde{p}_{2,1}
\end{equation}
where $\tilde{p}_{1,2}$ are defined as in~\ref{def:mom}, $s = -(1+M^2/Q^2)^2Q^2$ and $2p_1 \cdot p_2 = (1+M^4/Q^4)Q^2$.

\item Massless gauge bosons and on-shell massive fermions $p_1^2=p_2^2=m^2 \ll -s$ and 
\begin{equation}
p_{1,2}=\tilde{p}_{1,2}-\frac{m^2}{Q^2}\tilde{p}_{2,1}.
\end{equation}

\item Massless gauge bosons and on-shell fermions of two types, with a small and a large mass, $p_1^2=M^2, \, p_2^2=m^2$ and $q^2=0$, $m \ll M$.
\end{enumerate}

\subsubsection{Limit (a)}

In this limit the integral becomes,
\begin{eqnarray}
I_a(Q^2,m^2;d)=\int \frac{d^dk}{(k^2 -2 p_1 k)(k^2 -2 p_2 k)(k^2 -m^2)}
\end{eqnarray}
We define the threshold expansion parameter for this limit as $x = m^2/Q^2$, where $Q^2$ is the large scale.

The ${\cal U}$ and ${\cal F}$ polynomials in terms of the alpha parameters are respectively given by 
\begin{eqnarray}
{\cal U} = x_1+x_2+x_3 \\
{\cal F} =x_1\, x_2+x\, x_1\, x_3+x\, x_2\, x_3+x\, x_3^2
\end{eqnarray}
and the Gr{\"o}bner basis elements for the Landau equations that ${\cal F}$ satisfy are,
\begin{eqnarray}
\mathbb{G} = \left\lbrace (-1+x)\,x\, x_3, x_2 + x\, x_3, x_1 + x\, x_3 \right\rbrace
\end{eqnarray}
Clearly, we have only the Identity transformation : $\left\lbrace x_1\rightarrow x_1,x_2\rightarrow x_2,x_3\rightarrow x_3 \right\rbrace$ and the normal vectors we get in this limit are
\begin{equation}
\left(
\begin{array}{c|c}
Normal & Facet: top/bottom(1/-1) \\
\hline
 \lbrace 0,  0,  0 \rbrace & \lbrace -1 \rbrace\\
 \lbrace -1,-1,-1 \rbrace & \lbrace 1 \rbrace\\
 \lbrace 0 , 0 , -1\rbrace & \lbrace 1 \rbrace \\
 \lbrace -1 , 0,-1 \rbrace & \lbrace -1 \rbrace\\
 \lbrace 0 , -1 , -1 \rbrace & \lbrace -1 \rbrace\\
\end{array}
\right)
\end{equation}
which in turn correspond to the following regions:
\begin{enumerate}[label=\roman*)]
\item The hard region : $\lbrace 0, 0, 0 \rbrace$
\item The 1-collinear region: $\lbrace -1, 0, -1 \rbrace$
\item The 2-collinear region: $\lbrace 0, -1, -1 \rbrace$
\end{enumerate}
ASY also reports the same regions. In \cite{Smirnov_applied}, it is confirmed that only the hard and collinear contributions suffice for the evaluation of this diagram at leading order.
\subsubsection{Limit (b)}

In this limit, we have the leading order integral
\begin{equation}
 I_b(q^2,m^2;d)= \int \frac{d^d k}{k^2 \left(k^2-2 k p_1-M^2\right) \left(k^2-2 k p_2-M^2\right)}
 \end{equation}
The threshold expansion parameter is $x=M^2/Q^2$.

The Symanzik Polynomials and the Gr{\"o}bner basis for the Landau equation satisfied by ${\cal F}$ are
respectively,
\begin{equation}
 {\cal U}=x_1+x_2+x_3,
 \end{equation}
 \begin{equation}
 {\cal F}=x_1 x_2 x^2+2 x_1 x_2 x+x_1 x_3 x+x_2 x_3 x+x_1 x_2
\end{equation} 
The Gr{\"o}bner basis of Landau equations are,
\begin{equation}
\mathbb{G} = \left\{x^2 x_3,x_2 (x+1)^2+x x_3,x_3 \left(x_2+x x_3\right),x_1+x_2+2 x x_3\right\}
\end{equation}
Once again, we have only the Identity transformation and the normal vectors identified are
\begin{equation}
\left(
\begin{array}{c|c}
Normal & Facet: top/bottom(1/-1) \\
\hline

 \lbrace 0 , 0 , 0\rbrace  & \lbrace -1 \rbrace\\
 \lbrace -2,-2,-2\rbrace & \lbrace 1 \rbrace \\
 \lbrace -1 , 0 , -1 \rbrace & \lbrace -1 \rbrace\\
 \lbrace -1,-2,-1 \rbrace & \lbrace 1 \rbrace \\
 \lbrace 0 , -1 , -1\rbrace & \lbrace -1 \rbrace\\
 \lbrace -2 , -1 , -1 \rbrace & \lbrace 1 \rbrace\\
 \lbrace -1 , -1 , -2 \rbrace & \lbrace -1 \rbrace\\
 \lbrace -1 , -1 , 0 \rbrace & \lbrace 1 \rbrace\\
\end{array}
\right)
\label{eq:norm_sudakovb}
\end{equation}
and finally the distinct regions are that correspond to the scalings in Eq.~\ref{eq:norm_sudakovb} are:
\begin{enumerate}[label=\roman*)]
\item The hard region : $\lbrace 0, 0, 0 \rbrace$
\item The 1-collinear region: $\lbrace -1, 0, -1 \rbrace$
\item The 2-collinear region: $\lbrace 0, -1, -1 \rbrace$
\item The ultra-soft region: $ \lbrace -1, -1, -2 \rbrace $
\end{enumerate}
We see in this case also the complete agreement of our result with ASY and also with the contributions, reported in \cite{Smirnov_applied}. 

\subsubsection{Limit (c)}
For this limit, the integral is,
\begin{equation}
I_c(q^2,m^2;d)=\int \frac{d^d k}{k^2 \left(k^2-2 k p_1\right) \left(k^2-2 k p_2\right)}
\end{equation}
with the threshold expansion parameter $x = m^2/Q^2$.

The Symanzik Polynomials ${\cal U}$ and ${\cal F}$ are,
\begin{equation}
{\cal U}=x_1+x_2+x_3
\end{equation}
and
\begin{equation}
{\cal F}=x^2 x_1 x_2+x x_1^2+x x_2^2+x_1 x_2
\end{equation}
and the Gr{\"o}bner basis elements for the Landau equation satisfied by ${\cal F}$ is given by,
\begin{equation}
\mathbb{G} = \left\{\left(x^2-1\right)^2 x_2,2 x_1-x \left(x^2-3\right) x_2\right\}
\end{equation}
using which, we once again see that only Identity transformations are required. 

Finally, the distinct 
normal vectors corresponding to the facets are 
\begin{equation}
\left(
\begin{array}{c|c}
Normal & Facet: top/bottom(1/-1) \\
\hline
 \lbrace -1,-2,-1 \rbrace & \lbrace 1 \rbrace\\
 \lbrace -1 , 0 , -1 \rbrace & \lbrace -1 \rbrace\\
 \lbrace 0 , 0 , 0\rbrace  & \lbrace -1 \rbrace\\
 \lbrace -2,-2,-2 \rbrace & \lbrace 1 \rbrace\\
 \lbrace 0 , -1 , -1\rbrace  & \lbrace -1 \rbrace \\
 \lbrace -2 , -1 , -1 \rbrace & \lbrace 1 \rbrace\\
\end{array}
\right)
\end{equation}
 The regions are :
\begin{enumerate}[label=\roman*)]
\item The hard region : $\lbrace 0, 0, 0 \rbrace$
\item The 1-collinear region: $\lbrace -1, 0, -1 \rbrace$
\item The 2-collinear region: $\lbrace 0, -1, -1 \rbrace$
\end{enumerate}

\subsubsection{Limit (d)}
In this limit, the integral is 
\begin{equation}
I_d(q^2,m^2;d)=\int \frac{d^d k}{k^2 \left(k^2-2 k p_1\right) \left(k^2-2 k p_2\right)}
\end{equation}

The threshold expansion parameter here is $x = m^2/M^2$. The Symanzik polynomials are
\begin{eqnarray}
{\cal U}=x_1+x_2+x_3\\
{\cal F}=x_1^2+x\, x_1\, x_2+ x_1\, x_2+x\, x_2^2
\end{eqnarray}

The Gr{\"o}bner basis elements that generate the same ideal as the Landau equation for ${\cal F}$ is given
by, 
\begin{equation}
\mathbb{G} = \left\{(x-1)^2 x_2,2 x_1+(x+1) x_2\right\}.
\end{equation}
Once more with only the Identity transformation,
$\left\lbrace x_1\rightarrow x_1,x_2\rightarrow x_2,x_3\rightarrow x_3 \right\rbrace$ we the following
distinct normals,
\begin{equation}
\left(
\begin{array}{cc}
Normal & Facet: top/bottom(1/-1) \\
\hline
 \lbrace 0,-1,0 \rbrace & \lbrace 1 \rbrace\\
 \lbrace 0 , 0 , 0 \rbrace & \lbrace -1 \rbrace\\
 \lbrace -1 , -1 , -1 \rbrace & \lbrace 1 \rbrace\\
 \lbrace 0 , -1 , -1 \rbrace & \lbrace -1 \rbrace\\
\end{array}
\right)
\end{equation}
The scalings above correspond to the following regions:
\begin{enumerate}[label=\roman*)]
\item The hard region : $\lbrace 0, 0, 0 \rbrace$
\item The 1-collinear region: $\lbrace 0, -1, -1 \rbrace$
\end{enumerate}

\section{Discussion and Conclusion}
\label{sect:disc-conclusions}

The MoR is a powerful technique for obtaining expression of Feynman amplitudes at any order. The field however needs a robust algorithm for systematically finding all the regions. The study 
of such algorithms may provide insight into ideas crucial for validity of the MoR. In 
this work, we present an algorithm to identify the regions using ideas from power geometry, which is a 
powerful technique for analyzing properties of algebraic polynomials.

Our algorithm, ASPIRE, has allowed us to develop an
implementation in Mathematica where we have also used
external programs including UF.m and NDConvexHull.m. 
We have benchmarked the code by reproducing one and two loop examples from the literature. 
The salient steps in developing this algorithm can be summarised as follows:
\begin{enumerate}
\item We translate the traditional problem of finding the regions in the momentum and alpha parameter space to purely in the Alpha-parameter space by integrating out the loop momenta and then expanding the resulting integral in the new regions.

\item Using the form of the Landau Equations in the Alpha-parameter space, we reduce the problem to finding the leading order behavior of solution of the Landau equations in the different regions. Instead of working with 
${\cal F}$ or ${\cal U}$, which are homogeneous, we use the polynomial ${\cal G} = {\cal U} + {\cal F}$, as defined by~\cite{LeePomeransky}, which has
the advantage of not being homogeneous, while the truncated polynomials are quasi-homogeneous. 

\item These regions lie in the neighborhood of the origin and we systematically expand the integral in these neighborhoods via simple transformations which are obtained from the analysis of the Gr\"obner basis. These transformations are strongly constrained via the delta function in the parametric integral which forces $\Sigma \alpha_i = 1$, where $\alpha_i$ are the Alpha-parameters.

\item The expansion of the integrals in these neighborhoods is obtained via the use of an old but unvisited topic of \emph{Power Geometry}. We use only a small portion of this powerful technique to speed up our algorithm for finding the regions in the Alpha-parameter space. This technique allows one to obtain the leading order expansion of the integral in a particular neighborhood by analyzing the support of a ${\cal G}$ polynomial,  and constructing its Newton Polytope.

\item The regions are then defined as scalings of the Alpha-parameters which lead to the leading order expansion of the integral in that particular neighborhood of the origin.

\item The scalings are obtained by finding the normals to the surfaces of the Newton Polytope with a constraint on the first component of the normal vector.

\item We demonstrate that linear transformations in the Alpha-parameters lead to non-trivial transformations of the Newton Polytope, which leads to the uncovering of previously hidden regions.

\item We present a stand-alone documented Mathematica implementation of the above developed algorithm and provide several examples covering one loop and two loop amplitudes.
\end{enumerate}

      Our results are in agreement with previous work in this field by Jantzen, 
Smirnov~\cite{Jantzen:2012mw} and Pak and Smirnov~\cite{pak_smirnov}. We also settle the issue of 
\emph{PreResolve} as raised in the work of Jantzen, Smirnov and Smirnov~\cite{Jantzen:2012mw} and provide a 
Mathematically justified way of unveiling the full set of transformations that one needs, to successfully 
identify all the regions in the Alpha parameter space.

      Future extensions of this work includes looking into the connection between the sub-leading contributions of regions and the resulting geometry in the Alpha-parameter space as well as predicting the number of distinct regions a priori by studying the topology of amplitudes.  For the moment, this work, as indeed is the case with the work of Pak and Smirnov, Jantzen, Smirnov and Smirnov, is based on the ${\cal U}$ and ${\cal F}$ polynomials. Having more than one approach based on these may be profitable in the sense of eliminating the possibility of missing some regions when one or the other is used at one given time. The work here is based on the general properties of Landau singularities and closer to the spirit of the classical analyses of Feynman amplitudes to analyze their analyticity in the past, is now being employed to identify the regions associated with Feynman diagrams.
      
      The investigation of multi-loop non-planar vertex diagrams encountered in references,
 e.g.,~\cite{Fujimoto:1995ev} and also in the book of Smirnov~\cite{Smirnov_applied} are an interesting class 
 of amplitudes where our initial analysis suggests a rich family of solutions of the Gr\"obner Basis elements. Treatment of such non-trivial integrals would require automation of the analysis of Gr\"obner Basis elements for determining the complete set of transformations which will lead to the identification of all contributing regions.
 As a preliminary example we
solve Landau equations and the Gr\"obner basis of the non-planar vertex diagram considered
by \cite{Smirnov_applied} in the appendix ~\ref{sect:two loop non planar}.    It may be readily  seen that this is highly complicated
and we do not yet have the full solution for this system, and work is in progress.
 
 It may also be noted here that the present work, as in work of Pak and Smirnov, Jantzen, Smirnov and Smirnov is limited to dimensionally regularized
Feynman diagrams. We only provide the regions that maximally partition the Alpha parametric space at leading order. With the regions in hand one still needs, for a complete evaluation of the original integral at leading order, to use appropriate phase space regulators, as discussed in~\cite{Becher:2011dz}, and follow the procedure of Jantzen~\cite{Jantzen:2011nz} to perform zero-bin subtractions in the framework of dimensional regularization. In the past, analytic regulators above and
beyond dimensional regularization have been employed in the MoR approach to separate soft and collinear regions, see e.g.~\cite{Becher:2011dz}. The presence of the analytic regulators has been automated in ASY code as well as in FIESTA \cite{pak_smirnov,Smirnov:2015mct}.Such an extension to the ASPIRE algorithm would be very interesting. although beyond the scope of 
the current work.

An important question to ask is whether the ASPIRE algorithm can
find regions that the Asy algorithm cannot?  In order to answer
this question, we have visited some examples that have been
studied using ASY.  These include the diagrams found in refs\cite{KaiserSchweizer}
all of which have been analyzed by us using ASY,  
correspond to (a) the $J$ type one-loop
integral, (b) the fish diagram with the kaon loop, and
(c) the fish diagram with the $\pi-K$ loop.
When implemented on ASPIRE the results have been confirmed. In addition 
we have considered the two loop Master Integral vertex diagram
in eq. (7.30) of ref.~\cite{Smirnov_applied} and find agreement
between the results from ASY and ASPIRE. To this extent, at the
present level of investigation we find complete agreement.
It may yet be that ASPIRE has the potential to probe new regions in
the setting of non-planar diagrams not necessarily in the Sudakov
limit.  For the moment, this analysis has not been performed
partly due to the highly non-trivial Gr\"obner basis elements.
Such investigations are deferred to the future.

  All the work reported here has used the Alpha parametrization. While it has been convenient to analyze the regions here, the connection to the actual scaling behavior in the momentum space is less transparent. In order to actually assign the nature of the scalefulness to an isolated region, one has to go back to the momentum scaling behavior. This identification has been carried out by hand. This part of the algorithm needs to be automated as well.

\section*{Acknowledgements}
It is a pleasure to thank Thomas Becher for a reading of the manuscript
and several suggestions on the presentation.
We also thank V.A.Smirnov for useful correspondence on the subject.  BA is partly supported by the MSIL Chair of the Division of Physical and 
Mathematical Sciences, Indian Institute of Science. RS acknowledges the support of DST-INSPIRE Fellowship [IF150859]. SR wishes to thank Dishant Pancholi for very useful discussions and comments. 

\appendix
\section{Appendix}
\subsection{Gr{\"o}bner Basis}
Finding the Gr{\"o}bner basis of an ideal over a ring is a commonly occurring problem in computational algebra. It is used to study systems of algebraic equations.

\textbf{Definition:}
The Gr{\"o}bner basis $\mathbb{G}$ of an ideal $\mathcal{I}$ over a polynomial ring $\mathcal{R}$ is the generating set of $\mathcal{I}$ with respect to some monomial ordering with the property that the leading term of any polynomial in $\mathcal{I}$ is divisible by the leading term of some element in $\mathbb{G}$.

One of the most important properties of Gr{\"o}bner basis of an ideal containing a set of algebraic varieties is that the zeros shared by the system of equations are also shared by the Gr{\"o}bner basis elements.

In the example below we have given the Gr{\"o}bner basis for an arbitrary set of polynomials using the Buchberger's algorithm\cite{bberger}.

Consider the polynomials,\[p_1=x^3+y^3-2x^2y \quad p_2=x^2+y^2-3xy\]

The Gr{\"o}bner basis for the minimal ideal containing the two polynomials is \[\mathcal{I}=\langle y^4,x y^2,x^2-3 x y+y^2 \rangle\]

The two given polynomials have the common root $(x,y)=(0,0)$ which is immediately evident from the obtained Gr{\"o}bner basis.
\subsection{Rotation of Newton Polytope in Alpha-parametric space}
\label{sect:rotation}

The normal vectors of surfaces of Newton polytope are interpreted as the physical regions. In such a case, rotation of the Newton polytope in the alpha parametric space will yield a different set of normal vectors. However, the physical regions represented by both of them does not change since the boundary subset does not change. 

Consider a polynomial:
\begin{equation}
G(x,y) = x+y + 2 x^2 y + x y^2
\end{equation}
The support of the above polynomial is 
\begin{equation}
\label{eq:unrotated set}
S(g) = \lbrace (1,0), (0,1), (2,1), (1,2) \rbrace
\end{equation}
The Newton polytope for the above is: 

\begin{figure}[H]
\centering
\includegraphics[scale=0.5]{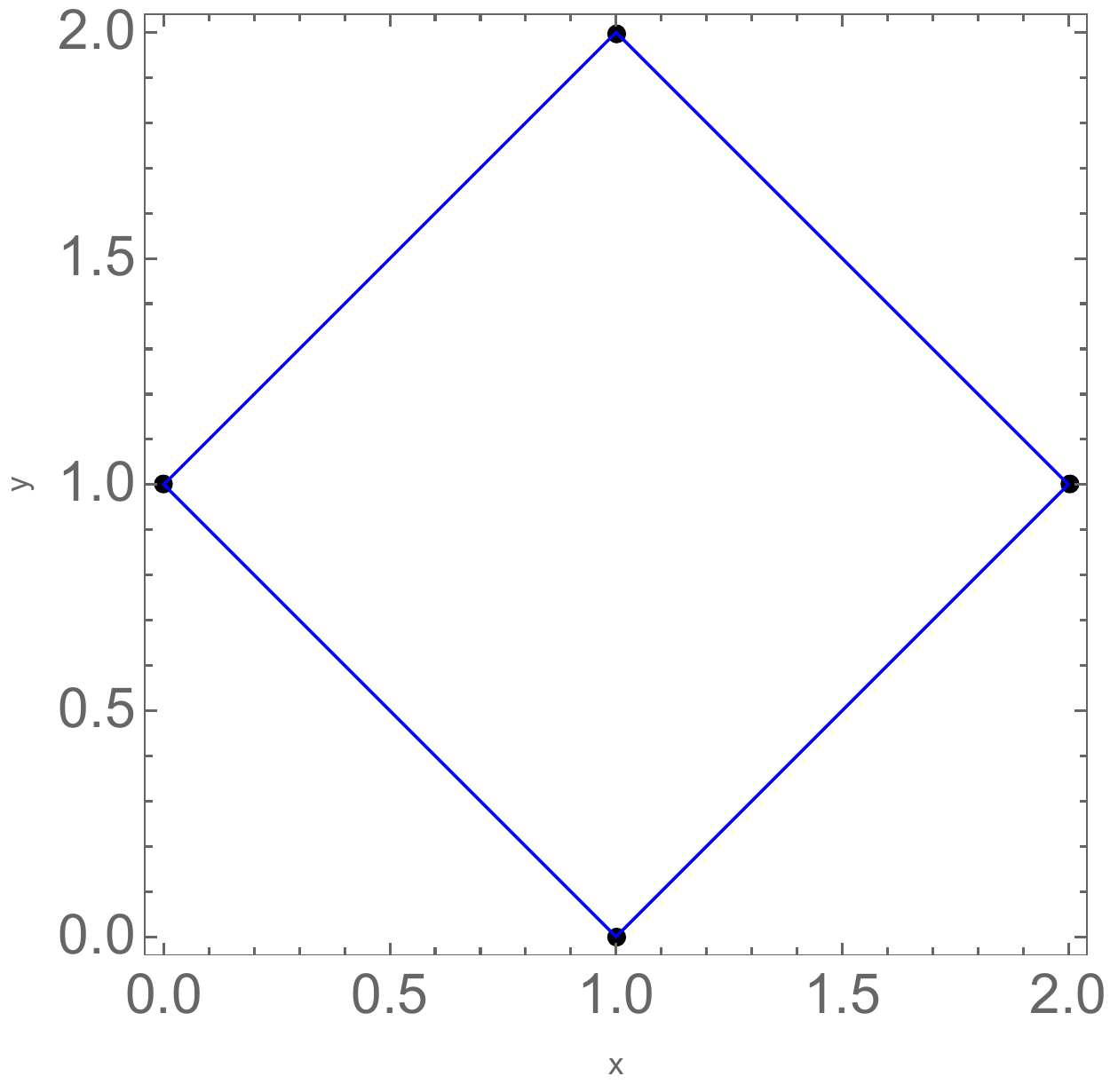}
\caption{Newton Polytope for ~\ref{eq:unrotated set}}
\end{figure}

Rotation of the Newton polytope corresponds to making a linear transformation of the support which translates to redefinition of the alpha parameters themselves. For the above let us make the transformation
\begin{equation}
\label{eq:transformation}
x \rightarrow x y^{\sqrt{3}} \quad  y \rightarrow x^{-\sqrt{3}} y
\end{equation}
In the above we performed the transformation

\begin{equation}
\begin{bmatrix}
1 && -\sqrt{3} \\
\sqrt{3} && 1
\end{bmatrix}
\begin{pmatrix}
1\\
0
\end{pmatrix}
=
\begin{pmatrix}
1\\
\sqrt{3}
\end{pmatrix}
\quad
\begin{bmatrix}
1 && -\sqrt{3} \\
\sqrt{3} && 1
\end{bmatrix}
\begin{pmatrix}
0\\
1
\end{pmatrix}
=
\begin{pmatrix}
-\sqrt{3}\\
1
\end{pmatrix}
\end{equation}

The new polynomial we get is 
\begin{equation}
G'(x,y) = 2 x^{2-\sqrt{3}} y^{2 \sqrt{3}+1}+x^{1-2 \sqrt{3}} y^{\sqrt{3}+2}+x^{-\sqrt{3}} y+x y^{\sqrt{3}}
\end{equation}

In the above it needs to be noted that the matrix is not orthogonal. The only condition that needs to be satisfied by a matrix, $M$ representing a rotation of the Newton polytope is $M^T M = c \mathtt{I}$, $c \in \mathtt{R}$ and $\mathtt{I}$ is the identity.

The new support of polynomial $G'(x,y)$ is 
\begin{equation}
S(g') = \lbrace (2-\sqrt{3},1+2 \sqrt{3}), (1-2\sqrt{3}, 2+\sqrt{3}), (-\sqrt{3},1), (1, \sqrt{3})  \rbrace
\end{equation}
The new Newton polytope for the new support is :
\begin{figure}[H]
\centering
\includegraphics[scale=0.5]{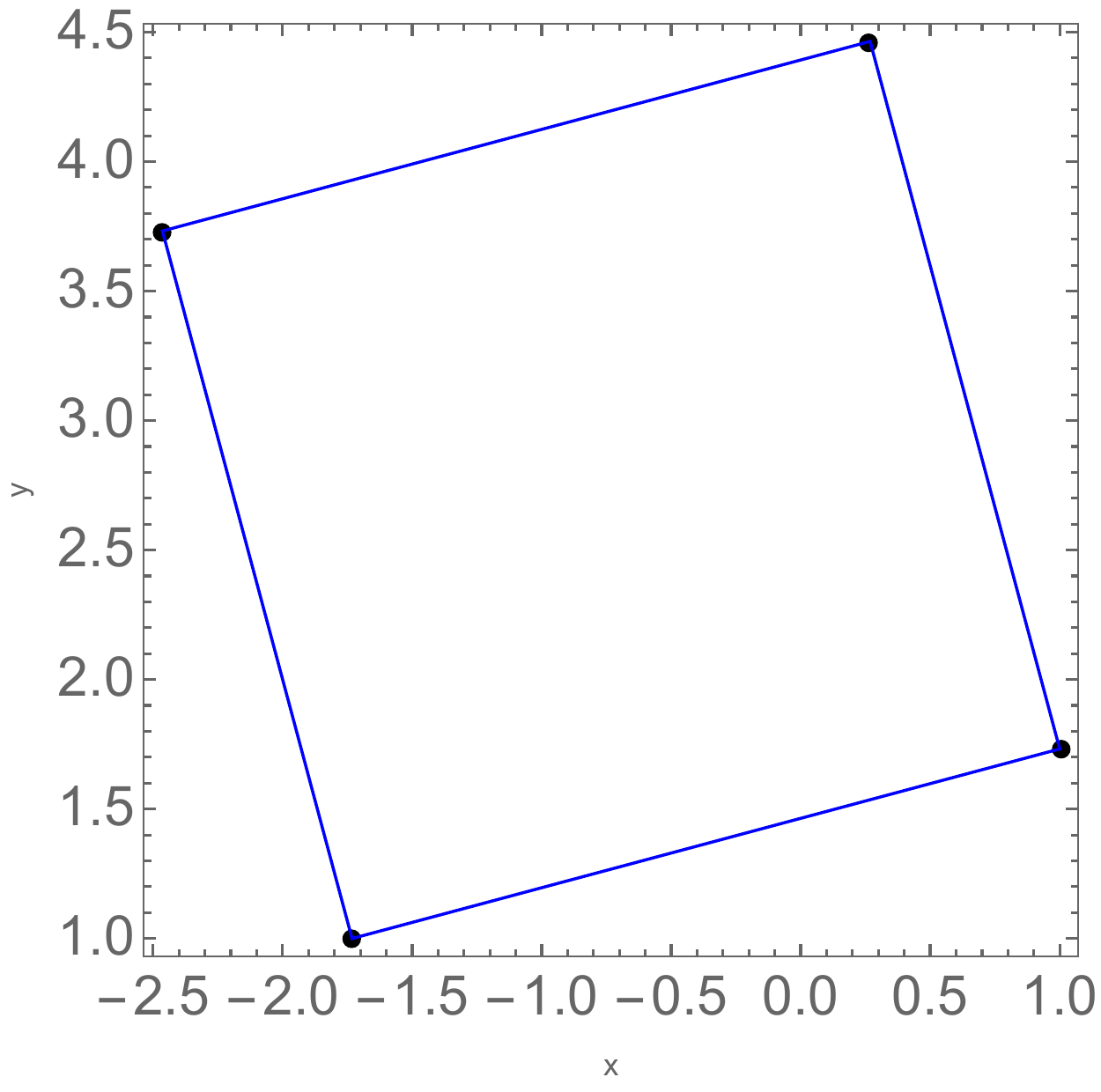}
\caption{Newton Polytope after transformations ~\ref{eq:transformation}}
\end{figure}

It is evident that rotation of the Newton polytope does change the normal vector of the surfaces. However, the boundary subset of the Newton polytope remains the same and thus the rotated and unrotated normal vectors would give the same boundary subset and hence the same truncated polynomial which implies that they both correspond to the same region. This can be seen Mathematically. Let us take the boundary subset of the Newton polytope $S_j^1$. Now for a normal vector $n$ we have,

\begin{equation}
\overrightarrow{n}.\overrightarrow{s} = k, \quad \forall \overrightarrow{s} \in S_j^1, k \in \mathtt{R}
\end{equation}
Rotating the polytope corresponds to 
\begin{equation}
\overrightarrow{s} \rightarrow M \overrightarrow{s} ,\quad \overrightarrow{n} \rightarrow M \overrightarrow{n}.
\end{equation}
Thus, we now have the condition,
\begin{equation}
\overrightarrow{n}.\overrightarrow{s} \rightarrow \overrightarrow{n} M^T M \overrightarrow{s} = c \overrightarrow{n}.\overrightarrow{s} = c k, \quad c \in \mathtt{R}
\end{equation}
Thus, we see that at the end of the transformation the boundary subset does not change. Hence the truncated polynomial still does not change and the truncated polynomial integrates to the same expression as the unrotated one. After making such a transformation, of course one needs to also include the jacobian in the integral and also change the limits of integration if necessary. One interesting thing to note here is that one can generate an infinite set of regions by rotating the polytope. These look different from each other at the outset but in fact represent the same process.
The purpose of this discussion is to show that the
invariances of the Newton polytope under the
rotations above, does not change the regions
identified in a substantive manner, in the case of the analysis of the U and F polynomials.  Rather the effect of the rotations is
to reexpress the rotated regions in terms of the original ones.
This does not give any further information on scaleless regions,
are required to be eliminated in any event.

\subsection{External Packages}
\subsubsection*{UF.m}
$\mathcal{UF}$ is a Mathematica based package designed for extracting the Symanzik polynomials, U and F, from the alpha representation of any multiloop Feynman integral. The code is openly available for redistribution \cite{UF}.\\
In our codes, $\mathcal{UF}$ function of the package \textit{UF.m} takes as input, for a particular Feynman integral, the set of loop momenta, set of all propagators and the set of kinematical substitutions as input. \\
Example:
\[UF[\lbrace k \rbrace,\lbrace -( k^2-m^2),-k^2 \rbrace,\lbrace m^2 \rightarrow x \rbrace],\] gives the output \[\lbrace x[1]+x[2],x x[1]^2+x x[1] x[2],1 \rbrace, \] where $x[1]$ and $x[2]$ are the Alpha-parameters. The output has three parts : first and second elements represent the ${\cal U}$ and ${\cal F}$ polynomials respectively and the third entry is the number of loops.

\subsubsection*{NDConvexHull.m}
This package includes implementations of several algorithms like Chan's algorithm, gift wrapping, quick hull, incremental convex hull, for finding the convex hull of a set of points in multi-dimensional space. They take a set of points as input and return a sorted list of vertex points and a sorted list of simplexes as output. This package was developed by Loren Petrich \cite{Petrich}.\\
In this work, we use the function \textbf{CHNQuickHull} for finding the convex hull of a set of points using the Quick Hull algorithm.\\
Example: Consider the following set of points \[\mathbb{P} = \lbrace (0,0,0), (1,0,0), (0,1,0), (0,0,1), (1,0,1), (1,1,0), (0,1,1)\rbrace. \]\\
CHNQuickHull[$\mathbb{P}$] gives the output in terms of the point ids which are the position of the points in $\mathbb{P}$.\\

\underline{Output:}

$\left\{\{1,2,3,4,5,6,7\},\left(
\begin{array}{ccc}
 1 & 2 & 6 \\
 1 & 3 & 4 \\
 1 & 4 & 2 \\
 1 & 6 & 3 \\
 2 & 4 & 5 \\
 2 & 5 & 6 \\
 3 & 6 & 7 \\
 3 & 7 & 4 \\
 4 & 7 & 5 \\
 5 & 7 & 6 \\
\end{array}
\right)\right\},$

\noindent
where the first list is the set of point ids of the given points and the second matrix is the surfaces of the Newton polytope in terms of those ids.

\subsection{Description of Mathematica Functions Implemented For This Work}
\subsubsection*{getMul}
We use this function for finding the support of a given polynomial. This function returns the vector exponents of the corresponding variables in a given monomial term.

\vspace{5pt}

Usage: \\
getMul[$x^2 y,\lbrace x,y\rbrace$] 

Output:
$\lbrace 2,1 \rbrace$

\subsubsection*{getNormal}
This function computes the components of the normal vector of a plane given the set of points lying on the plane and a set of points lying below the plane.

\vspace{5pt}

Usage:\\
getNormal[$\lbrace \lbrace 1,2,0 \rbrace, \lbrace 2,1,0  \rbrace,\lbrace 1,0,2 \rbrace \rbrace,\lbrace \lbrace 1,0,1 \rbrace, \lbrace 1,1,0 \rbrace, \lbrace 0,1,1 \rbrace \rbrace $]

\vspace{2pt}

 Output: $\lbrace v[1] \rightarrow 1,v[2] \rightarrow 1, c \rightarrow 3 \rbrace$, \\

where the first matrix of the input of the function represents the points on the surface and the second matrix contains the points lying below the surface. The output are the components of the normal vector of the surface with the zeroth component as 1.

\subsubsection*{genNormalCoordinates}
This function takes as input the set of points on a facet of the Newton Polytope. It finds the set of points below the surface using the points that lie on the surface and then uses the getNormal function for getting the normal vector.

\vspace{5pt}

Usage:\\
genNormalCoordinates[$\lbrace$ points on surface of Newton polytope $\rbrace$, $\lbrace$ set of all points $\rbrace$]

Output: Components of the normal of the facet

\subsubsection*{UniqueRegions}\label{unqReg}
It takes as input the set of normal vectors and the length of the normal vectors, removes any instance of \textbf{Null} from the set and then eliminates all normals related by a constant shift. This gives us a unique set of regions. 

\vspace{5pt}

Usage:\\
UniqueRegions[$\lbrace \lbrace v[1] \rightarrow 0,v[2] \rightarrow 0 \rbrace, Null, \lbrace v[1] \rightarrow 1, v[2] \rightarrow 0 \rbrace, Null, \lbrace v[1] \rightarrow 1, v[2] \rightarrow 1 \rbrace \rbrace$]

Output: $\lbrace \lbrace 0,0 \rbrace,\lbrace 1,0 \rbrace \rbrace $
\subsection*{Scalecheck}
This function checks for scaleless integrals. It takes as input the leading order U and F polynomial in a particular region and then checks if the polynomials are proportional to themselves under rescaling of a subset of alpha parameters. A region is scaleless if:

\begin{equation}
F_{lead.}[ \lbrace \alpha_i \rbrace] \propto F_{lead}[c \lbrace \alpha_j \rbrace \cup \lbrace \alpha_k \rbrace ], 
\end{equation}
where $  \lbrace \alpha_j \rbrace \subset  \lbrace \alpha_i \rbrace, \lbrace \alpha_j \rbrace \cup \lbrace\alpha_k\rbrace = \lbrace \alpha_i \rbrace$ and $ c \in \mathbb{R}$ 

\vspace{5pt}

Usage:\\
Scalecheck[$x+y$, $x^2 y z$, $\lbrace$x,y,z$\rbrace$]

\vspace{2pt}

Output: Scaleless

\subsection{Preliminary Analysis of two loop Non-Planar Diagram in the Sudakov Limit}
\label{sect:two loop non planar}
We perform initial analysis of a two loop non-planar vertex diagram studied in \cite{Smirnov_applied} in limit ~\ref{lim_a}. The integral is
\begin{equation}
I_{NP}=\int \int \frac{d^d k d^d l}{[(k+l)^2-2p_1.(k+l)][(k+l^2-2p_2.(k+l))](k^2-2p_1.k)(l^2-2p_2.l)(k^2-m^2)(l^2-m^2)}
\end{equation}

The Gr\"obner Basis of the Landau equations for this integral contains 22 terms each having multiple solution branches. Exploring all the branches needs to be automated. For the purposes of demonstration, we list some of the Gr{\"o}bner Basis elements :

\begin{equation}
\begin{split}
\mathbb{G} & = \{(x-1) x \left(x_4-x_5\right) x_6 \left(x_4+x_6\right) \left(x_5+x_6\right), (x-1) x \left(x_4-x_5\right) \left(x_4+x_6\right) \left(x_5^2-x_6^2\right),\\ & (x-1) x \left(x_4-x_5\right) \left(x_4+x_6\right) \left((8 (x-1) x+1) x_4-4 (x-1) x \left(x_5-x_6\right)+x_6\right),\\ & (x-1) x \left(x_4-x_5\right) x_6 \left(x_4+x_6\right) \left(x_4+x_5+2 x_6\right), \left(x_4-x_5\right) \left(x_4+x_6\right) \left(x_5+x_6\right) \left(x_4+x \left(x_5+x_6\right)\right),
\\ & \left(x_4-x_5\right) \left(x_4+x_6\right) \left(x_4^2+\left(x_5+2 x_6\right) x_4+x x_6 \left(x_5+x_6\right)\right), \cdots \}
\end{split}
\end{equation}

From the above list, let us analyze the Gr\"obner Basis element:
\[(x-1) x \left(x_4-x_5\right) \left(x_4+x_6\right) \left((8 (x-1) x+1) x_4-4 (x-1) x \left(x_5-x_6\right)+x_6\right)\]

In the above element the factor $(x-1) x \left(x_4-x_5\right) \left(x_4+x_6\right)$, becomes zero if $x_4-x_5=0$ or $x_4+x_6=0$. The positivity of the Alpha-parameters implies that $x_4+x_6=0$ gives $x_4=x_6=0$. Thus, this solution branch corresponds to a trivial transformation. 

The factor $\left((8 (x-1) x+1) x_4-4 (x-1) x \left(x_5-x_6\right)+x_6\right)$, can be mapped to zero if $x_5-x_6 = 0$, $x_4 = 0$ and $x_6 = 0$. However, this implies that $x_4=x_5=x_6=0$. This once again results in a trivial transformation. Hence, the only non-trivial transformation that one needs to perform here is to map $x_4-x_5 = 0$. 

More complicated Gr\"obner Basis elements have multiple solution branches. All of these branches look distinct at the beginning but as one does a careful study of these, it might be revealed that these seemingly distinct branches lead to similar transformations. To find the complete set of transformations needed to detect all the regions an automated approach to analyze and identify all the distinct solution branches is required and is deferred to future versions of our algorithm.

While analyzing this system in our scheme by considering only the Identity transformation (i.e. the transformation which leaves the Alpha parameters unaltered), we get the same regions as obtained from ASY. However, the complicated nature of the Gr{\"o}bner bases may allow for other non-trivial transformations which can be studied in the future. A notebook for the preliminary study has been provided.

\subsection{List of Mathematica Notebooks Used in This Work}
We provide brief description of the Mathematica notebooks used in this discussion. It may be noted that care has been taken to produce the U polynomial with the correct positive sign denoted by the extension "P" at the end of the name of the notebook.

\begin{tabular}{|c|c|}
\hline 
{\bf File} & {\bf Explanation} \\ 
\hline 
\bf{TwoPointOneLoopP.nb} & Contains the demonstration of the algorithm for \\ & obtaining the regions associated with\\ & the integral representing\\ & a two point one loop diagram. \\ 
\hline 
\bf{FivePointOneLoopP.nb} & Regions in five point one loop diagram have been shown.  \\ 
\hline 
\bf{PionFishP.nb} & Reveals the regions in a two loop Fish diagram having\\ & internal pion loop in the context of $\pi$ - K scattering. \\ 
\hline 
\bf{ScalarTriangleDiagramLimit(a)P.nb} & Regions in Scalar Triangle diagram in\\ & Sudakov limit-(a) have been obtained in this notebook. \\
\hline
\bf{ScalarTriangleDiagramLimit(b)P.nb} & Regions in Scalar Triangle diagram in \\ & Sudakov limit-(b) have been obtained in this notebook. \\ 
\hline
\bf{ScalarTriangleDiagramLimit(c)P.nb} & Regions in Scalar Triangle diagram in\\ & Sudakov limit-(c) have been obtained in this notebook. \\ 
\hline
\bf{ScalarTriangleDiagramLimit(d)P.nb} & Regions in Scalar Triangle diagram in\\ & Sudakov limit-(d) have been obtained in this notebook. \\ 
\hline 
\bf{SudakovNonPlanarP.nb} & Non-planar diagram in the Sudakov limit\\ & with idenentity  transformation \\ & only has been studied. \\
\hline
\end{tabular}

\end{document}